\documentclass[aps,prd,twocolumn,groupedaddress,nofootinbib]{revtex4-2}

\RequirePackage[colorlinks,citecolor=blue,urlcolor=blue]{hyperref}

\def\simlt{\stackrel{<}{{}_\sim}}

\usepackage{soul} 

\usepackage{amsthm,amsmath}
\RequirePackage[OT1]{fontenc}
\usepackage{enumerate}
\usepackage{aasmacros}
\usepackage{graphicx}
\usepackage{amsfonts}
\usepackage{bm,bbm}
\usepackage[]{algorithm2e}
\usepackage{epstopdf}
\usepackage[font=small,skip=0pt]{caption}
\usepackage[font=small,skip=0pt]{subcaption}
\RequirePackage{color}
\usepackage[dvipsnames]{xcolor}
\usepackage{multirow}
\usepackage[normalem]{ulem}

\graphicspath{{./}{figs/}{../}}

\newcommand{\figref}[1]{Fig.~\ref{#1}}
\RequirePackage{bm}

\usepackage[font = {it}]{caption}

\newcommand{\R}{\reels}

\newcommand{\Msun}{M_{\odot}}

\renewcommand{\Re}{\mathbb{R}}
\renewcommand{\R}{\Re}

\newcommand{\diag}[1]{\mathbf{#1}}

\begin{document}


\title{Differentiating small-scale subhalo distributions in CDM and WDM models using persistent homology}


\author{Jessi Cisewski-Kehe}
\email[]{jjkehe@wisc.edu}
\affiliation{Department of Statistics, University of Wisconsin-Madison, 1300 University Ave, Madison, WI 53706, USA}
\author{Brittany Terese Fasy}
\affiliation{Gianforte School of Computing, Montana State University, 357 Barnard Hall, P.O. Box 173880, Bozeman, MT,  59717-3880, USA}
\author{Wojciech Hellwing}
\affiliation{Center for Theoretical Physics, Polish Academy of Sciences, Al. Lotników 32/46, 02-668 Warsaw, Poland}
\author{Mark R. Lovell}
\affiliation{Science Institute, University of Iceland, Dunhaga 5, 107 Reykjav\'ik}
\author{Paweł Drozda}
\affiliation{Center for Theoretical Physics, Polish Academy of Sciences, Al. Lotników 32/46, 02-668 Warsaw, Poland}
\author{Mike Wu}
\affiliation{Computer Science Department, Stanford University, 353 Jane Stanford Way, Stanford, CA, 94305, USA}

\date{\today}

\begin{abstract}

The spatial distribution of galaxies at sufficiently small scales will encode information about the identity of the dark matter. We develop a novel description of the halo distribution using persistent homology summaries, in which collections of points are decomposed into clusters, loops and voids. We apply these methods, together with a set of hypothesis tests, to dark matter haloes in MW-analog environment regions of the cold dark matter (CDM) and warm dark matter (WDM) Copernicus Complexio $N$-body cosmological simulations. The results of the hypothesis tests find statistically significant differences (p-values $\leq$ 0.001) between the CDM and WDM structures, and the functional summaries of persistence diagrams detect differences at scales that are distinct from the comparison spatial point process functional summaries considered (including the two-point correlation function). The differences between the models are driven most strongly at filtration scales $\sim100$~kpc, where CDM generates larger numbers of unconnected halo clusters while WDM instead generates loops. This study was conducted on dark matter haloes generally; future work will involve applying the same methods to realistic galaxy catalogues.
\end{abstract}


\maketitle

\section{Introduction}
\label{sec:intro}

The large scale structure (LSS)---as defined by the spatial distribution of 
galaxies---encodes information on many vital aspects of the standard model of cosmology that remain open questions in physics \citep{davis1985evolution,bull2016beyond,bullock2017small,perivolaropoulos2021challenges}.
For example, the LSS is sensitive to the characteristics of dark energy, the unexplained phenomenon that drives the accelerated expansion of the Universe \citep{vdWeygaert11,Sanchez:2012aa} and also holds clues as to the nature of dark matter (DM). Typical LSS observables that are relevant for DM studies include the abundance of low mass galaxies \citep{Papastergis11,Kennedy14}, the paucity of galaxies in voids \cite{Tikhonov:2009uq} and the spatial distribution of MW satellite galaxies \cite{Lovell21}. An additional, as yet largely untapped, method for analysing LSS models is the application of topological methods to the distribution of galaxies and haloes. These methods describe the spatial distribution of points as different dimensional holes with clusters, filaments loops, and voids in dimensions 0, 1, and 2, respectively, and it is possible to envisage that the imprint of DM physics on the primordial density field may be detectable in their topological statistics \cite{vdWeygaert11,Watts2017}. 
In this paper we will apply topological methods in order to identify differences between two competing DM models. 
The simplest viable model of DM is the cold DM matter model (CDM), in which the DM particle has a negligible velocity dispersion at early times and thus DM halos are able to start collapsing early and in large quantities. The combination of CDM with the cosmological constant model of dark energy is known as $\Lambda$CDM.
This model has enjoyed success in  predicting the properties of the cosmic microwave background (CMB) radiation \cite{PlanckCP14} and the distribution of galaxies at large scales ($>$2~Mpc) \cite{Eisenstein05}. However, at smaller scales ($<$1~Mpc) there are tensions among others with the densities of dwarf galaxies that may hint at problems for the CDM model  \citep{bullock2017small}. Given the simultaneous failure to detect the particle physics candidates that correspond to CDM in direct detection experiments \cite{Akerib17,Aprile18} or in indirect detection observations \cite{Albert17}, it is important to consider alternatives.

\begin{figure*}[htp!]
  \centering
    \begin{subfigure}{.40\textwidth}
        \centering
        \includegraphics[width=\linewidth]{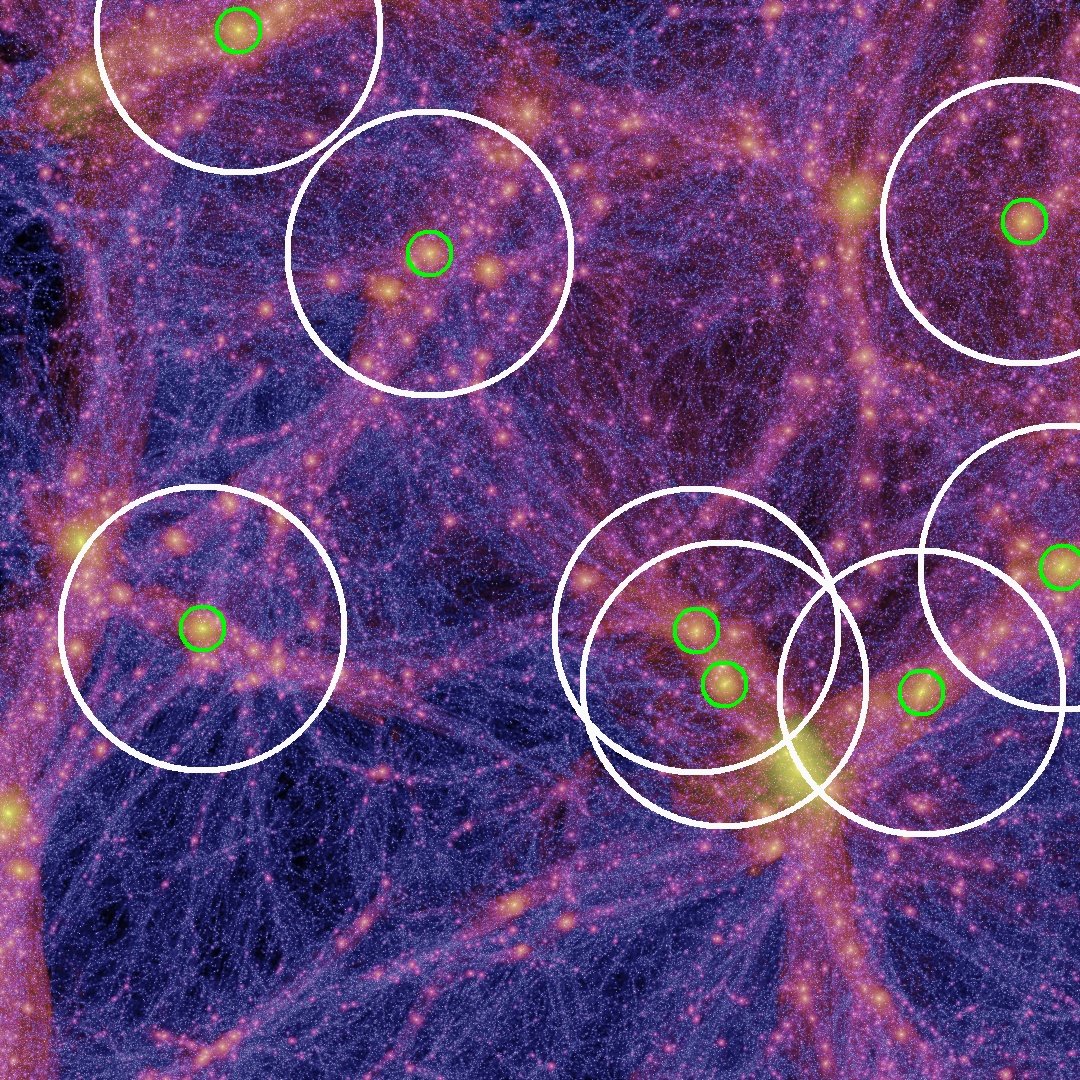}
        \caption{COCO-CDM}
        \label{fig:introDataCDM}
    \end{subfigure}
    \begin{subfigure}{.40\textwidth}
        \centering
        \includegraphics[width=\linewidth]{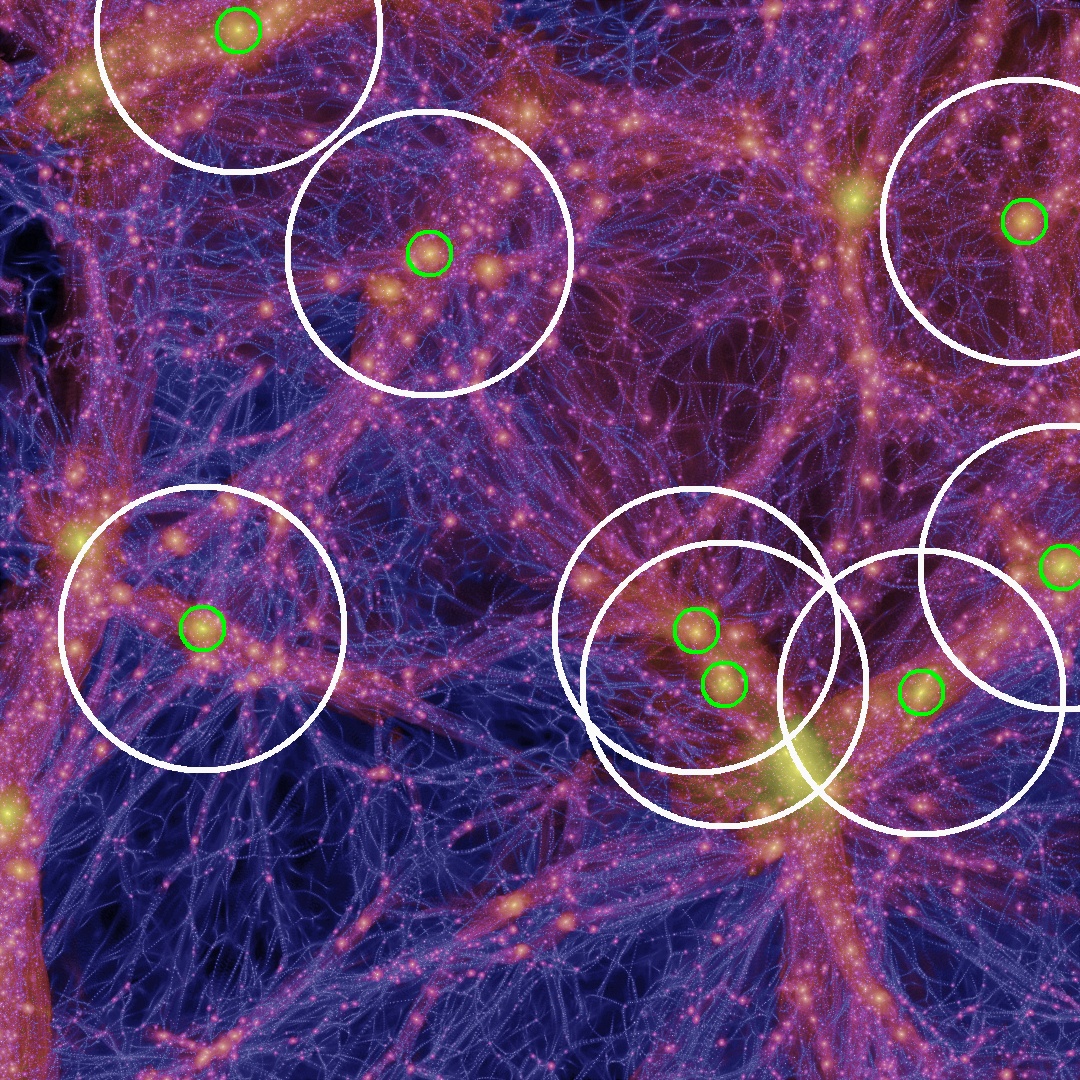}
        \caption{COCO-WDM}
        \label{fig:introDataWDM}
    \end{subfigure}
    \caption{Illustration of the DM distributions in the COCO-CDM (a) and COCO-WDM (b) simulations. Each image is a slice through the simulation of 23 Mpc on a side with an image depth of 10 Mpc. The image intensity encodes the DM column density and the image color indicates velocity dispersion. Eight of the 77 volumes used in this study (see \S\ref{sec:mw}) are included in this slice, and their locations are indicated as follows. The MW-analog halo on which each volume is centered is enclosed by a green circle, and the full extent of the analysis volume (a radius 3 Mpc) is shown with a white circle. Note that the apparent overlap of the white circles in this projection does not imply that the volumes overlap: there can still be considerable separation between the volumes in the depth direction.   See \S\ref{sec:coco} for details on the COCO data.
    }
    \label{fig:introData}
\end{figure*}

One compelling alternative to CDM is the warm DM (WDM) model, in which the DM particles have a significant velocity dispersion in the early Universe \citep{colombi1996large}.  The effects of this velocity dispersion include a drastic reduction in the number of low mass DM halos. In this study we compare simulations of these two models to determine whether persistent homology can detect differences in the DM halo spatial distribution.
In \figref{fig:introData}, we present images of two realizations of the Copernicus Complexio (COCO) cosmological volume \citep{Hellwing:2016tv}, one simulated with the CDM model, and the second with the WDM model \citep{bose2016copernicus}. The large scale distribution of matter is nearly identical in the two
images---thus WDM preserves the large scale successes in explaining the distribution of massive galaxies of
CDM---but at smaller scales the abundance of WDM halos is strongly suppressed relative to CDM, and the distribution of the remaining subhalos is much less homogeneous.

In this work, we investigate differences in the spatial distribution of DM haloes as described in CDM and WDM. The primary goal is to ascertain whether topological methods are sensitive to differences between the models and the second goal is to interpret the differences to determine whether topological methods have the potential to discern which DM model most accurately describes the properties of our own Universe. We restrict our analysis to the distribution of haloes, which will work as a proof of concept.  A comprehensive comparison with observations will require a mock galaxy catalogue and we defer this step to future work.

The persistent homology formulation of topology offers a novel way to represent, visualize, and interpret complex data by extracting homological features, which can be used to infer properties of the underlying structures. Homological features include the decomposition of halo distributions into clusters, filaments loops, and voids at different scales controlled by a parameter that is analogous to halo linking 
lengths---which in statistics is known as a filtration parameter---and persistent homology in particular tracks how the number of such features changes as the filtration parameter is increased. It has been successfully applied to problems in astronomy (e.g., Refs. \cite{Sousbie2011, SousbieEtAl2011, vdWeygaert11,cisewski2014non, Pranav:2017vy, Green:2019uz, Pranav:2019ws, Xu:2019ue, Cole:2020ul}), along with other areas of science (e.g., Refs. Ref. \cite{duong2012closed, bendich2014persistent, Lawson:2019tg, Berry:2020ws}).
 There have been proposals for hypothesis testing using persistent homology (e.g., Refs. \cite{Robinson:2017vm, bubenik2015statistical, Biscio:2019ub, Berry:2020ws,Krebs:2021vw}), which we build on as we construct tests that can detect differences between DM model predictions in the LSS.

We investigate several test statistics to discriminate between the CDM and WDM halo spatial distributions that are based on persistent homology functional summaries.  Each functional summary is a different transformation of information to a function that approximates a property of the topological features, and is a function of the filtration parameter.  We also consider different visualizations in order to investigate detected differences.

This paper is organised as follows. We begin with background on the cosmological simulation data we use in the analysis (\S\ref{sec:coco}), then we introduce persistent homology and functional summaries of persistence diagrams that are used in the proposed test statistics (\S\ref{sec:tda}).  Then the hypothesis testing framework is presented (\S\ref{sec:methods}), followed by the investigation of the cosmological simulation data (\S\ref{sec:coco_analysis}).
We end with concluding remarks (\S\ref{sec:conc}).

\section{Cosmological Simulation Data} \label{sec:coco}


This section begins with a description of the COCO simulations, and then continues with our procedure for selecting MW halo-analog sample regions.


\subsection{The Copernicus Complexio (COCO) cosmological simulations}

The COCO simulation volume constitutes a high resolution spherical region of space with a comoving radius of approximately 25~Mpc; the full (low-resolution) simulation volume is a periodic box 100~Mpc on a side.\footnote{All distances are in comoving Mpc.}
The numerical integration of the gravitational forces begins at redshift 127. The cosmological parameters are consistent with the 7-year results from the WMAP satellites: matter density $\Omega_0 = 0.272$, dark energy density $\Omega_{\Lambda} = 0.728$, $\Omega_b=0.04455$, Hubble parameter $h_0 = 0.704$, spectral index $n_s=0.967$, and power spectrum normalization $\sigma_8=0.81$.  
The mass of the simulation particle is $1.135 \times 10^5$ $M_{\odot}$. 
DM halos and subhalos were identified using the SUBFIND algorithm \citep{springel2001populating}, and the smallest permitted number of particles to identfy a subhalo is 20 particles. Our definition of halo mass is the total mass bound gravitationally to each halo as determined by the halo finder.

Two copies of this volume were run, the first applying CDM \cite{Hellwing:2016tv} and the second WDM \cite{bose2016copernicus}. Both simulations use the same initial phases, and differ in that the WDM simulation had wave amplitudes rescaled using the transfer function of a 3.3~keV thermal relic DM particle, with the relic mass chosen to be in agreement with the Lyman-$\alpha$ forest constraints of Ref.~\cite{viel2013warm}. This results in the suppression of structure on the scale of dwarf galaxies. These large-scale structure similarities between the WDM and CDM data due to the same initial phases are shown in \figref{fig:introData}. One issue peculiar to WDM simulations is the spurious numerical fragmentation of filaments into halos; these so-called spurious subhalos are identified and removed from the halo catalog using the algorithm described in Ref.~\cite{lovell2014properties}.

\subsection{Milky Way-analog DM halos and their associated halo samples} \label{sec:mw}


Given that we intend to use future work to compare the models with observations of galaxies around our own MW, we identify MW-analog halos and their surrounding regions in the two simulations. The criteria for our MW-analog halos were that they must be located within 21~Mpc of the center of the simulation\footnote{The central high-resolution sphere of COCO extends out to about 25 Mpc.}
and have a mass in the range $[0.5,2]\times10^{12}$~$\Msun$.  We also required that there be no other halo with a mass greater than $0.5\times10^{12}$~$\Msun$ within 0.7~Mpc. This procedure resulted in 77 MW-analog DM halos in each of the COCO CDM and WDM simulations; for each MW-analog DM halo in the WDM data, there is a matching MW-analog DM halo in the CDM.

\begin{figure*}[htp!]
  \centering
    \begin{subfigure}{.40\textwidth}
        \centering
        \includegraphics[width=\linewidth]{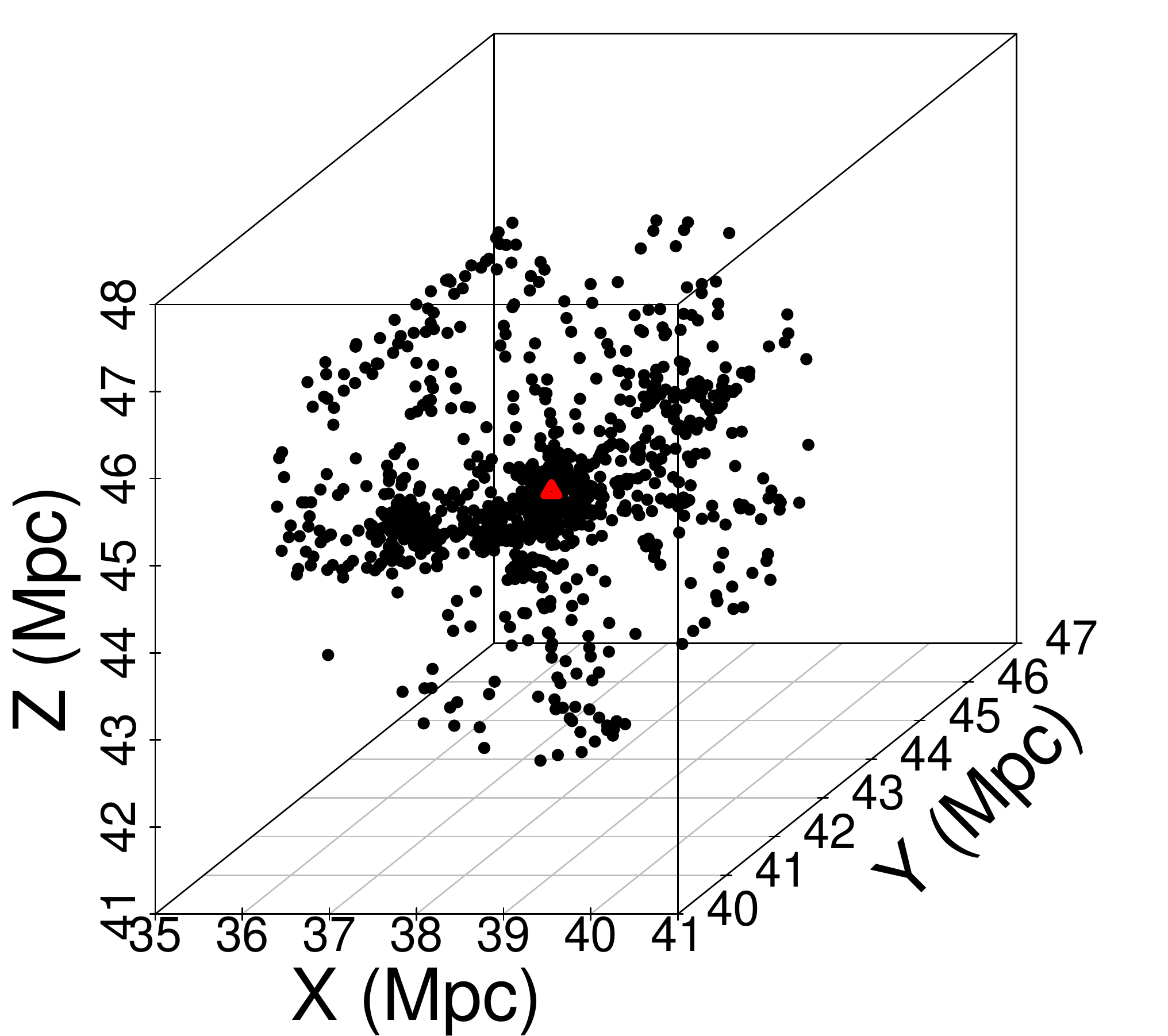}
        \caption{CDM}
        \label{fig:mw_cdm}
    \end{subfigure}
    \begin{subfigure}{.40\textwidth}
        \centering
        \includegraphics[width=\linewidth]{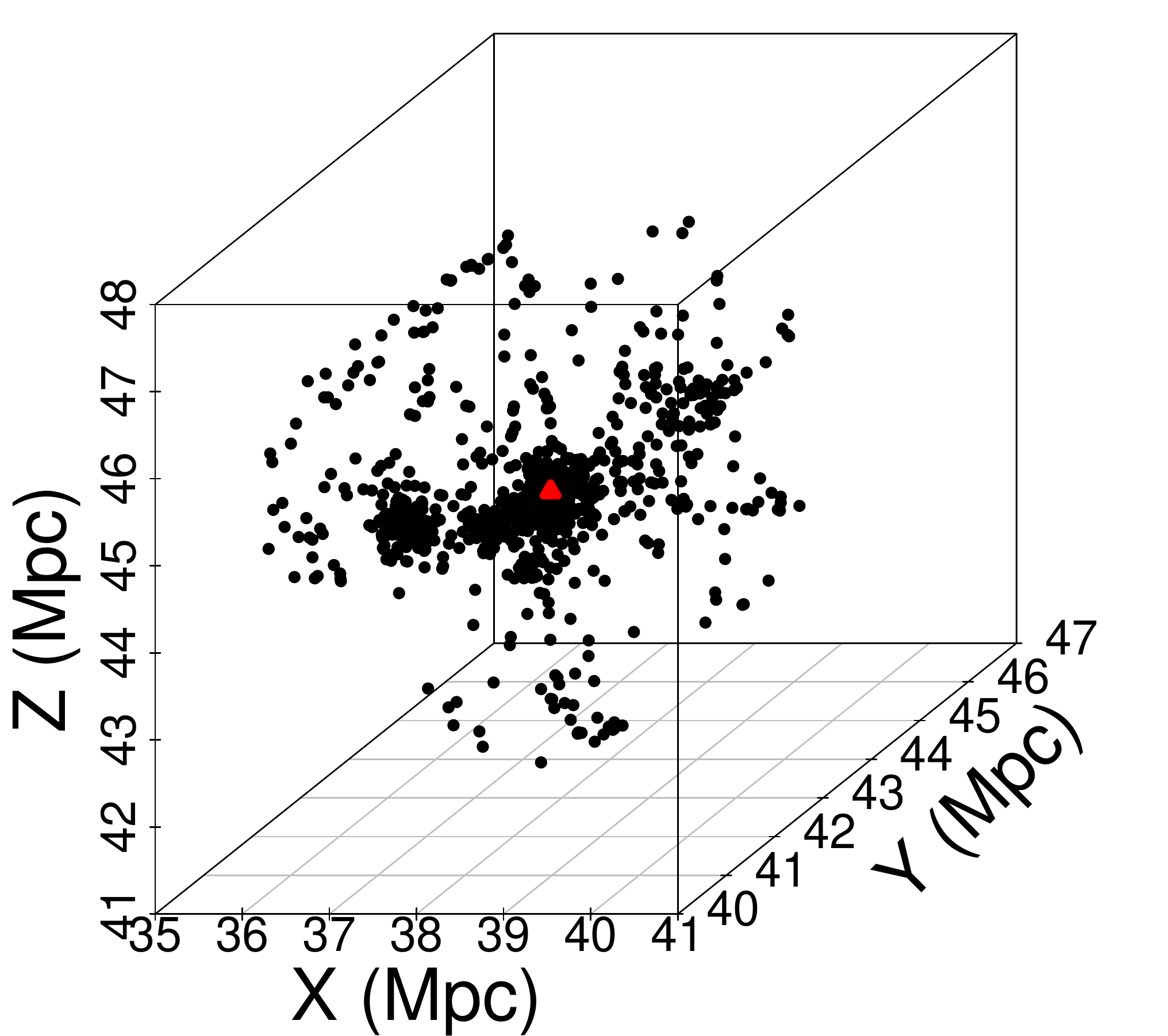}
        \caption{WDM}
        \label{fig:mw_wdm}
    \end{subfigure}
    \caption{A MW-analog DM halo neighborhood sample for (a) the COCO-CDM data and (b) the COCO-WDM data.  The red triangles indicate the MW-analog DM halo that was selected along with the other DM halos that are within 3 Mpc from it.}
    \label{fig:mw_example}
\end{figure*}

We now discuss our selection of halos in the vicinity of the 77 MW halo-analogs. In both CDM and WDM realizations we identify halos that are within 3~Mpc of the MW-analog\footnote{Objects with $\sim$0.2~Mpc of each analog are typically referred to as `subhalos' that orbit within the analog `host halo.' In this study we refer to all bound DM objects simply as `halos' and include all of them in our analysis, not drawing any distinction between `subhalos' and other classes of object.}. For the WDM case we include all halos in the 3~Mpc region. However, CDM forms hundreds of times more halos than WDM in our resolved mass range. If we were to include all CDM halos, the abundance difference would dominate our statistical results.
 Therefore, the CDM samples were downsampled to match the number of DM halos in the corresponding WDM sample. The downsampling was accomplished by selecting the most massive DM halos from each of the CDM samples.
An example of one of the MW-analog DM halo neighborhoods from COCO-CDM and COCO-WDM is displayed in \figref{fig:mw_example}.

The two sets of samples for the CDM and WDM data are defined as
\begin{equation}
\mathbb Y_c = \{\mathbf Y_{c,1}, \ldots, \mathbf Y_{c,77}\},
\mathbb Y_w = \{\mathbf Y_{w,1}, \ldots, \mathbf Y_{w,77}\}
\label{eq:samples}
\end{equation}
where $\mathbb Y_c$ and $\mathbb Y_w$ represent the set of 77 CDM and 77 WDM samples, respectively.  Each $\mathbf Y_{k,i} \in \mathbb R^{n_i \times 3}$ for $k = c,w$ and $i = 1, \ldots, 77$ where $n_i$ indicates the number of DM halos in sample $i$; an individual DM halo in simulation $k$ of sample $i$ is indicated by $Y_{k,i,j}$ for $j = 1, \ldots, n_i$.


\section{Topological Data Analysis Methods for Quantifying LSS} \label{sec:tda}
Homology is one way to study the features of topological spaces (e.g., manifolds),
specifically the dimensional ``holes''  in the space (e.g., connected components,
loops, voids). Persistent homology studies the spatial structure of a
parameterized family of topological spaces that keeps track of the so-called
\emph{births} and \emph{deaths} of homological features as a topological space
changes with a filtration parameter.  
In particular, we focus on point cloud data, where each point can represent some
unit of mass or an object (e.g., a point may represent a center of a DM halo).
In this section, we provide a brief overview of the necessary concepts; however, see, e.g.,
Refs.~\cite{munkres1984elements, hatcher2002algebraic, edelsbrunner2010computational} for a more thorough introduction to
algebraic and computational topology.
%
The homological features that are tracked in the filtration have cosmological
interpretations in dimensions zero, one, and two.  Before providing more details
about persistent homology, we explain the interpretation of different dimensional
holes with respect to the distribution of DM halos.

\paragraph{Clusters}
A \emph{connected component}, or zeroth-dimensional homology feature ($H_0$), is
a maximal subspace of a topological space that cannot be covered by two disjoint
open sets; that is, a connected component is a \textit{whole piece} of the
space.  
For example, under some assumptions on the topological space, a connected component is a cluster of data points.\footnote{There is no clear or established relationship between
the definition of homological clusters and galaxy clusters. In this paper, the
term `cluster' is only used for the homology definition.} In cosmology, the
connected components represent clusters of halos or galaxies. Persistent homology tracks
the appearance of new connected components and the merging of distinct
components.

\paragraph{Filaments and Loops}
A \emph{loop}, or one-dimensional homology feature ($H_1$), provides information
about the connectivity of data. As the filtration parameter increases, nearby
connected components can merge together in such a way that a loop is formed. 
For DM halos, this would appear as filaments of halos
joined together in a loop.

\paragraph{Cosmological Voids}
A \emph{void}, or two-dimensional homology feature ($H_2$), represents the
boundary of three-dimensional empty regions within the topological space (e.g.,
the boundary of a football). 
In cosmology,
these are the thin walls surrounding the low-density regions that are typically referred to as
cosmological voids.

\subsection{Persistent homology}

Persistent homology is a framework for computing the homology of a data set at
different scales.
Given a data set, one defines a filtration (that is, a sequence
of nested topological spaces) of intermediate
structures, on which the homology is computed at different values of the
filtration parameter. Homology group generators (e.g., $H_0$, $H_1$, $H_2$
features) are tracked as they form and die as the filtration parameter changes.
Various methods can be used in order to transform a discrete point set into a
connected topological space. For example, simplicial complexes (see below) such as the
Vietoris--Rips complex (VR complex) can be used, or a function can be defined
over the domain of the data using an empirical distance function or a kernel
density estimate (KDE) of the point cloud.
In this work, we use a VR complex to construct the filtration (discussed below).
Next, we introduce some key components of persistent homology.

\paragraph{Simplicial complexes}
A geometric $k$-simplex is the convex hull of $k+1$ affinely independent points.
For our 3D halo data, the simplices we use are zero-simplices (vertices),
one-simplices (edges), two-simplices (triangles), and three-simplices (tetrahedrons).
A face of a simplex is another simplex obtained by removing zero or more
points (e.g., a triangle has seven faces: itself, three edges,
and the three vertices).
A
simplicial complex, $\mathcal K$, is defined as a finite set of simplices such
that (i) if $\sigma \in \mathcal K$, then every face of $\sigma$ is also in
$\mathcal K$, and (ii) if $\sigma_1, \sigma_2 \in \mathcal K$, then either~$\sigma_1 \cap \sigma_2 \in \mathcal K$ or $\sigma_1 \cap \sigma_2 = \emptyset$.
In this work, we use simplicial complexes to represent topological spaces, as
they are the standard input to code to compute homology.

\paragraph{Filtrations}
A filtration is a sequence of nested topological spaces.  Given dataset~$y_1,
y_2, \ldots, y_n \in \mathcal Y \subseteq \mathbb R^3$, one common way to create
a simplicial complex is to choose some $t \in \R$ such that $t \geq 0$
and replace each $y_i \in \mathcal{Y}$ with a ball of
diameter $t$.  The Vietoris--Rips complex at scale $t$ (the $t$-VR complex) is created by representing each of
these balls as a vertex, and creating a $k$-simplex anytime there are $k+1$
balls that pairwise intersect.
Specifically:
\begin{equation}
VR_{t}(S) = \{\sigma \subseteq S \mid d(x,z) \leq t, \forall x, z \in \sigma,\}  \label{eq:vr}
\end{equation}
where $d(\cdot, \cdot)$ is the Euclidean distance \citep{edelsbrunner2010computational,
zomorodian2010fast}.
That is, $VR_{t}(S)$ is a simplicial complex containing
the vertex set $\mathcal S$, edges between all the vertices that are separated
by at most $t$, and triangles for sets of three vertices that have pairwise
distances of at most $t$.

We obtain the
VR filtration by
increasing $t$ from $0$ to $\infty$
(here, $t$ is referred to as the \textit{filtration parameter}).\footnote{In practice,
the maximum filtration value $t$ we consider correspond to the largest scales encompassed by a given galaxy/halo catalog.}
Note that $VR_{t_1}(S)$ is a subset of $VR_{t_2}(S)$ (i.e., $VR_{t_1}(S) \subseteq
VR_{t_2}(S)$) for $t_1 \leq t_2$.
Sometimes, for the right selection of $t$ and a dense enough sample,
we can recover the homology of $\mathcal{Y}$
(see, e.g.,~Ref. \cite{niyogi2008finding});
however, using the whole sequence
of complexes, we can recover information about~$\mathcal{Y}$
with more relaxed sampling conditions.

\begin{figure*}
  \begin{subfigure}{.3\linewidth}
    \centering
        \includegraphics[width=\linewidth]{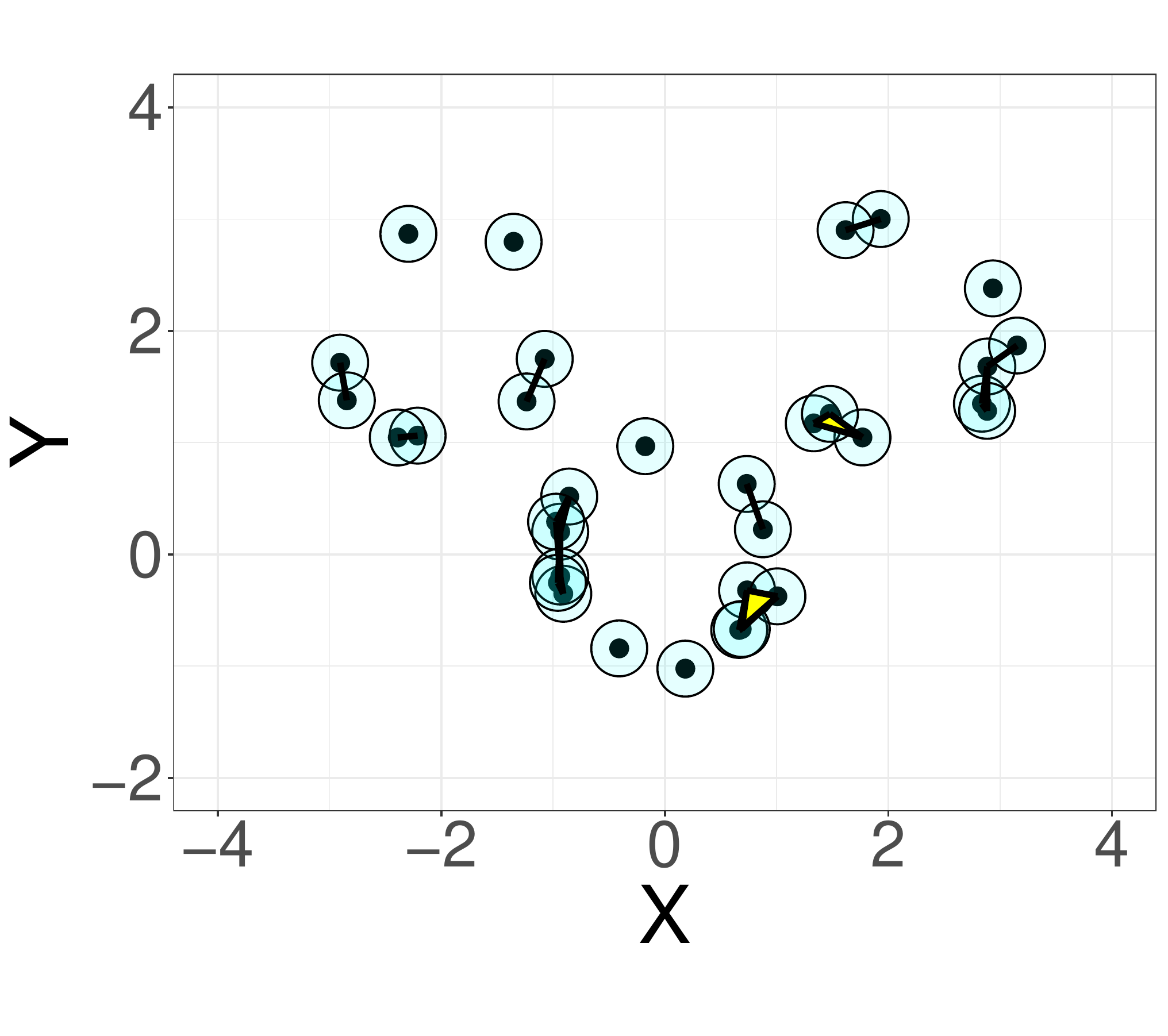}
    \caption{Data with $t=0.5$ balls}
    \label{fig:circle_1}
  \end{subfigure}
    \begin{subfigure}{.3\linewidth}
    \centering
        \includegraphics[width=\linewidth]{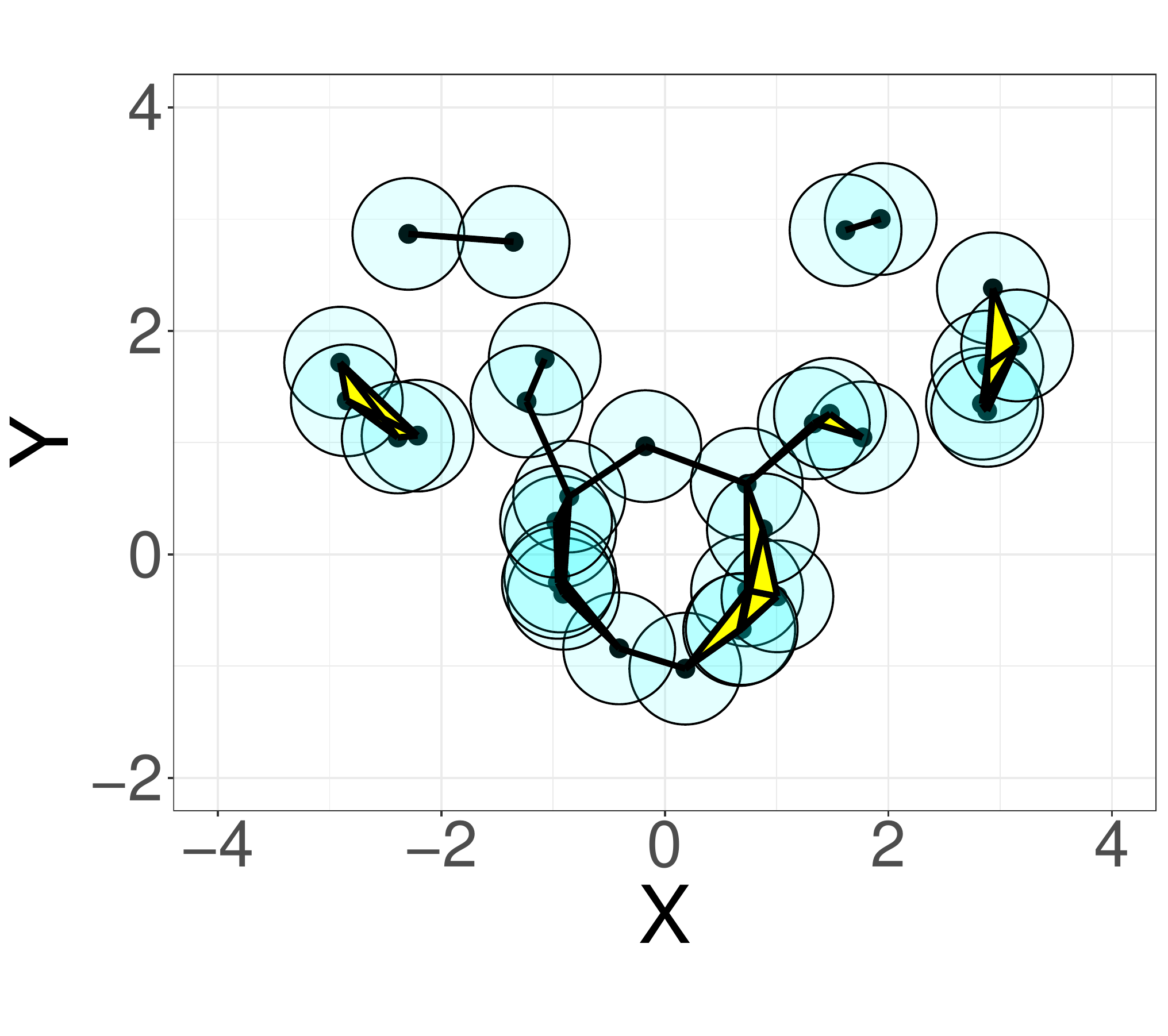}
    \caption{Data with $t=1$ balls}
    \label{fig:circle_2}
  \end{subfigure}
    \begin{subfigure}{.3\linewidth}
    \centering
        \includegraphics[width=\linewidth]{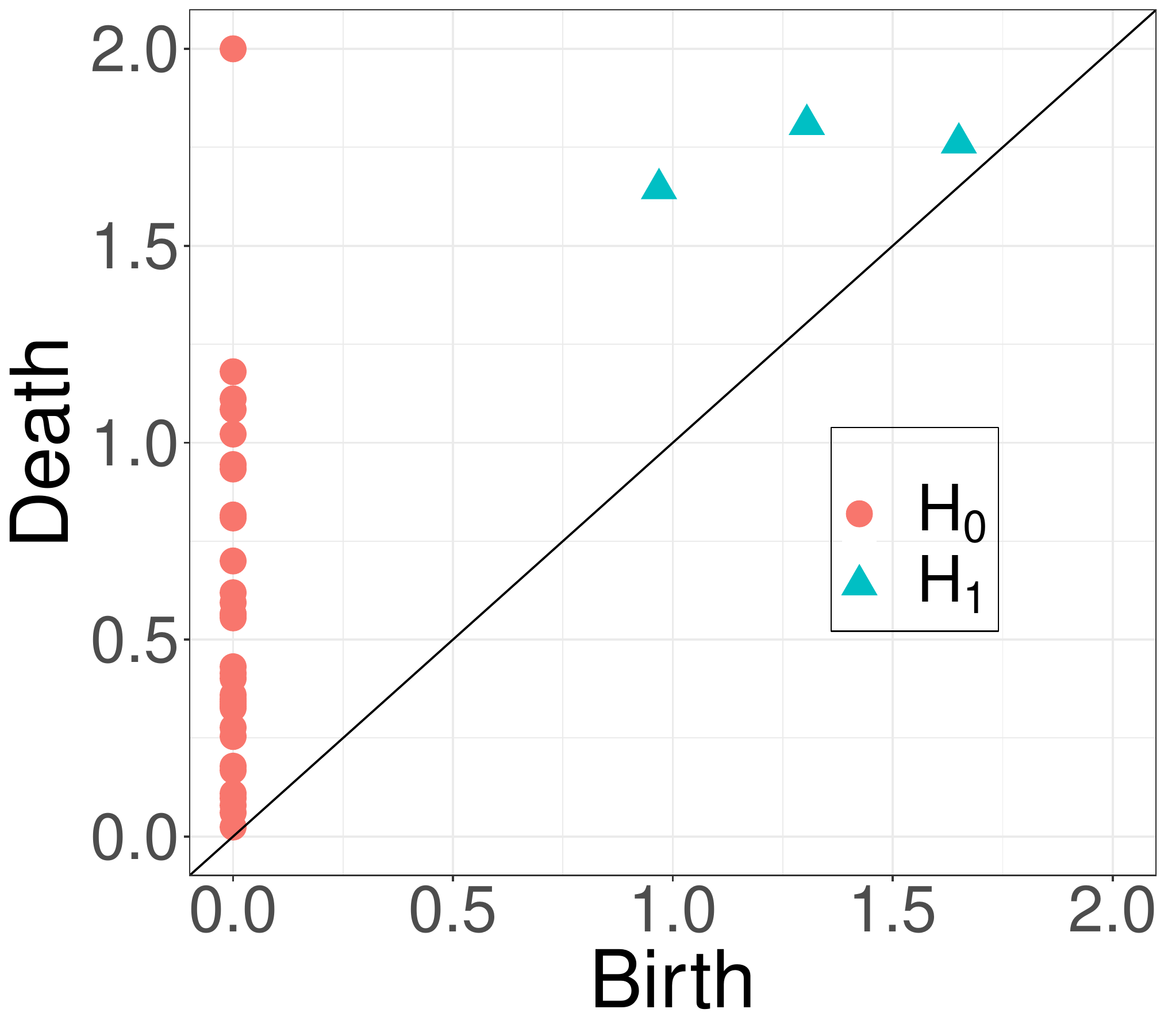}
    \caption{Persistence diagram}
    \label{fig:circle_diagram}
  \end{subfigure}
    \caption{Persistence diagram example where observations were sampled around
    three circles with noise. The data are displayed in (a) and (b) as black
    points (zero-simplices) with cyan balls with diameters of 0.5 and 1,
    respectively, along with the one- and two-simplices of the corresponding VR
    complexes.  The persistence diagram for the VR filtration of the points is
    displayed in (c) with the the three circles indicated by the cyan triangles
    ($H_1$).  The $H_0$ features represent the connected components, which all
    have birth times at 0.  } \label{fig:homologyexample}
\end{figure*}

To derive the persistent homology for a VR filtration, the homology of
$VR_{t}(S)$ is computed as $t$ changes.
If $t$ is initialized at $0$, then only the data points contribute to the
homology.
The evolving topological space is characterized by its homology as~$t$ increases
toward $\infty$.
For a $\mathcal Y \subseteq \mathbb R^3$, the persistent homology would then
track the connected components ($H_{0}$), loops ($H_{1}$), and voids ($H_{2}$)
that appear and disappear in the VR filtration.
An example of a VR filtration with a 2-dimensional domain is presented in
\figref{fig:homologyexample}: Figures~\ref{fig:circle_1} and \ref{fig:circle_2}
display the data points with balls of diameter t = 0.5 and 1, respectively,
along with the one- and two-simplices of the corresponding VR complex.
\figref{fig:circle_diagram} shows the persistence diagram for the data points
using the VR filtration, which is discussed next.

\paragraph{Tracking Homology Generators}

The birth and death times of the homology group generators are displayed in a
persistence diagram.  These times correspond to values of the
filtration parameter, which is the diameter of the balls $t$ when considering a
VR filtration.  Suppose a filtration is defined over some data points $y_1,
y_2, \ldots, y_n \in \mathcal Y \subseteq \mathbb R^3$, then a persistence
diagram, $\diag D$, can be written as a multiset of points:
\begin{equation}
    \diag D = \{(r_j, b_j, d_j): j = 1, \ldots, |\diag D|\} \cup \Delta
\end{equation}
where $(r_j, b_j, d_j)$ are the homology group dimension, the birth time, and
the death time, respectively, of feature $j$, $|\diag D|$ indicates the number
of homology group generators with~$d_j > b_j$, and $\Delta$ represents a set of 
points on the diagonal (birth time = death time) with infinite multiplicity.
The persistence diagram is a nice summary because small changes in the input
data~$\mathcal{Y}$ will result in only small changes in the diagram~\cite{cohen2007stability,chazal2016structure}.

Figures~\ref{fig:circle_1} and ~\ref{fig:circle_2}  show an example where the
filtration parameter, $t$, increases from~0.5 to~1. In that interval, the
homology changed from having 15 connected components ($H_0$'s) and zero loops
($H_1$'s) to having 5 connected component and 1 loop. The time in the filtration
when homology features appear, the \emph{birth} of the feature, and the time
when a feature joins other features, the \emph{death} of the feature, are
captured in a persistence diagram.  \figref{fig:circle_diagram} displays the
persistence diagram, where the location of each point represents the birth
height (x-axis) and death height (y-axis) of a homological feature for a VR
filtration, and the shape and color represent the homology group dimension.
A point $(\cdot,x,x)$ on the diagonal represents a feature with a zero-length
lifespan.  The \emph{persistence} of a point $(\cdot, b,d)$ is the length of the
interval of the persistence parameter that supports that feature: $|d - b|$. In
the persistence diagram, the distance from $(\cdot, b,d)$ to the diagonal is
proportional to this value; in fact, the (Euclidean) distance to the diagonal is
$\frac{|d-b|}{\sqrt{2}}$.
%
When working with empirical experimental or observational data it is necessary to be concerned
with the associated intrinsic noise of such measurements. This is especially important in the context
of a spatial distribution of objects derived either from N-body simulations or galaxy catalogs.
N-body simulations are limited by their spatial resolution, where their Monte Carlo sampling
nature starts to breaks down and is overrun by the shot-noise. The astronomical observations are limited by imperfections including those related to involved instruments which contribute to measurements errors.
For these reasons it is important in a persistent homology analysis to be able to distinguish between real features really present in the target and those that are noise-induced transients. 
A notion of a topological significance can be derived in this context by considering features with longer lifetimes as more significant, and those with short lifetimes (i.e., closer to the diagonal) as topological noise \citep{fasy2014confidence}.
Distinguishing between topological signal and noise is a problem of a great interest 
in real applications of TDA (e.g., Ref. \cite{Xu:2019ue}).

\subsection{Persistence diagram summaries} \label{sec:summaries}
While persistence diagrams and their individual features provide useful information about the topology of a data set, persistence diagrams are not easy objects to work with directly for statistical analyses. 
For example, the distance between two persistence diagrams can be calculated
using metrics such as the bottleneck distance or the $p$-Wasserstein distance,
but both are computationally expensive because they require finding a certain
optimal matching between the features on each diagram; see Equation~\eqref{eq:bottleneck} in the Appendix for the definition of the bottleneck distance.
Fr\'echet means and medians have been defined for spaces of persistence diagrams
\citep{Turner:2014kq}, but are also computationally expensive and not
necessarily unique (although ways around this exist as addressed in
Ref. \cite{Munch:2015fk}).  Instead, we consider transformations and summaries of
persistence diagrams that make computations more tractable \citep{Berry:2020ws}.
Below are several approaches that transform a persistence diagram into a
functional summary, which are used in \S \ref{sec:methods} to formulate test
statistics for hypothesis tests.

\begin{figure}[htp!]
  \begin{subfigure}{.75\linewidth}
    \centering
        \includegraphics[width=\linewidth]{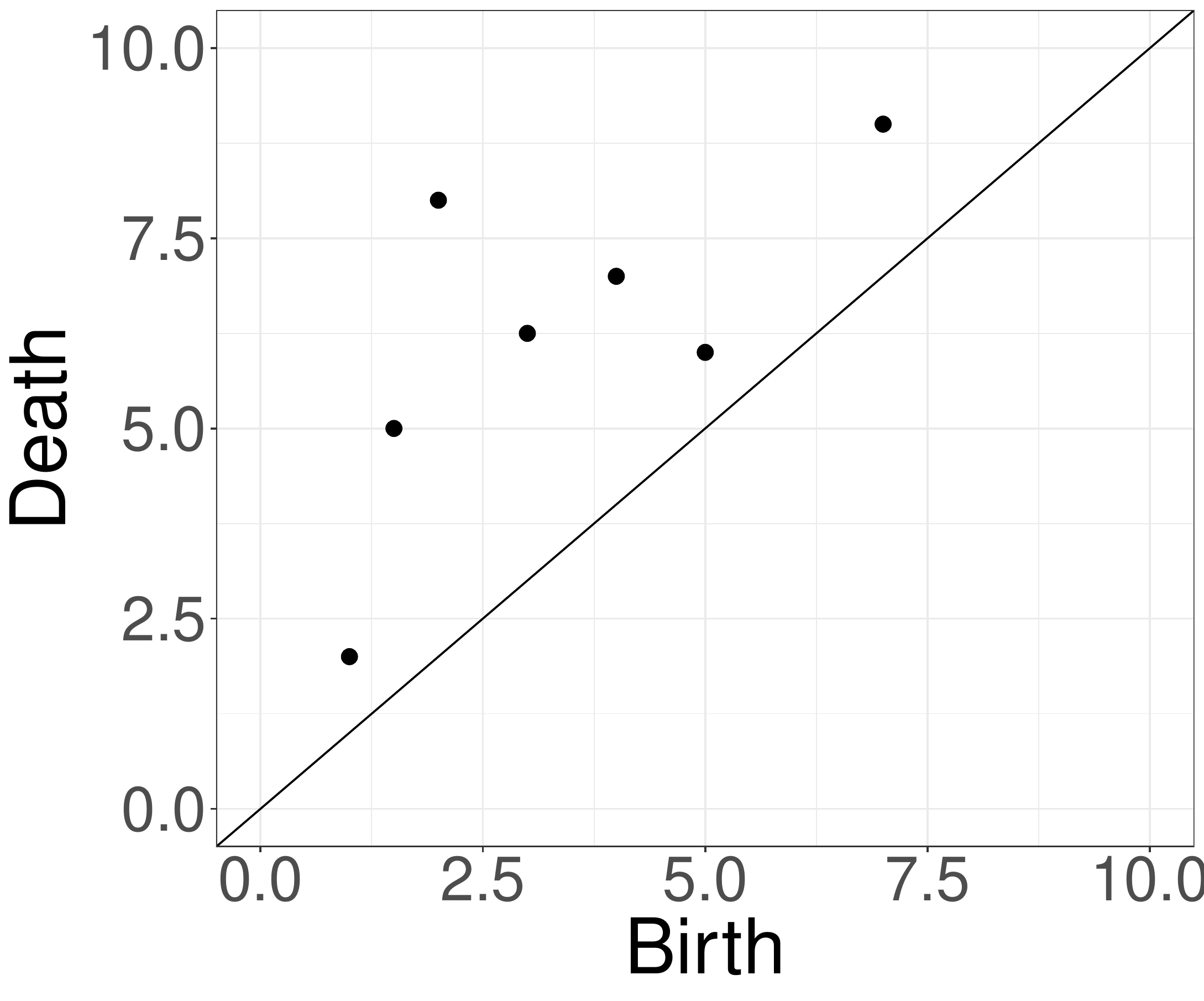}
    \caption{Persistence Diagram}
    \label{fig:landscape_diagram}
  \end{subfigure}
  
    \begin{subfigure}{.67\linewidth}
    \centering
        \includegraphics[width=\linewidth]{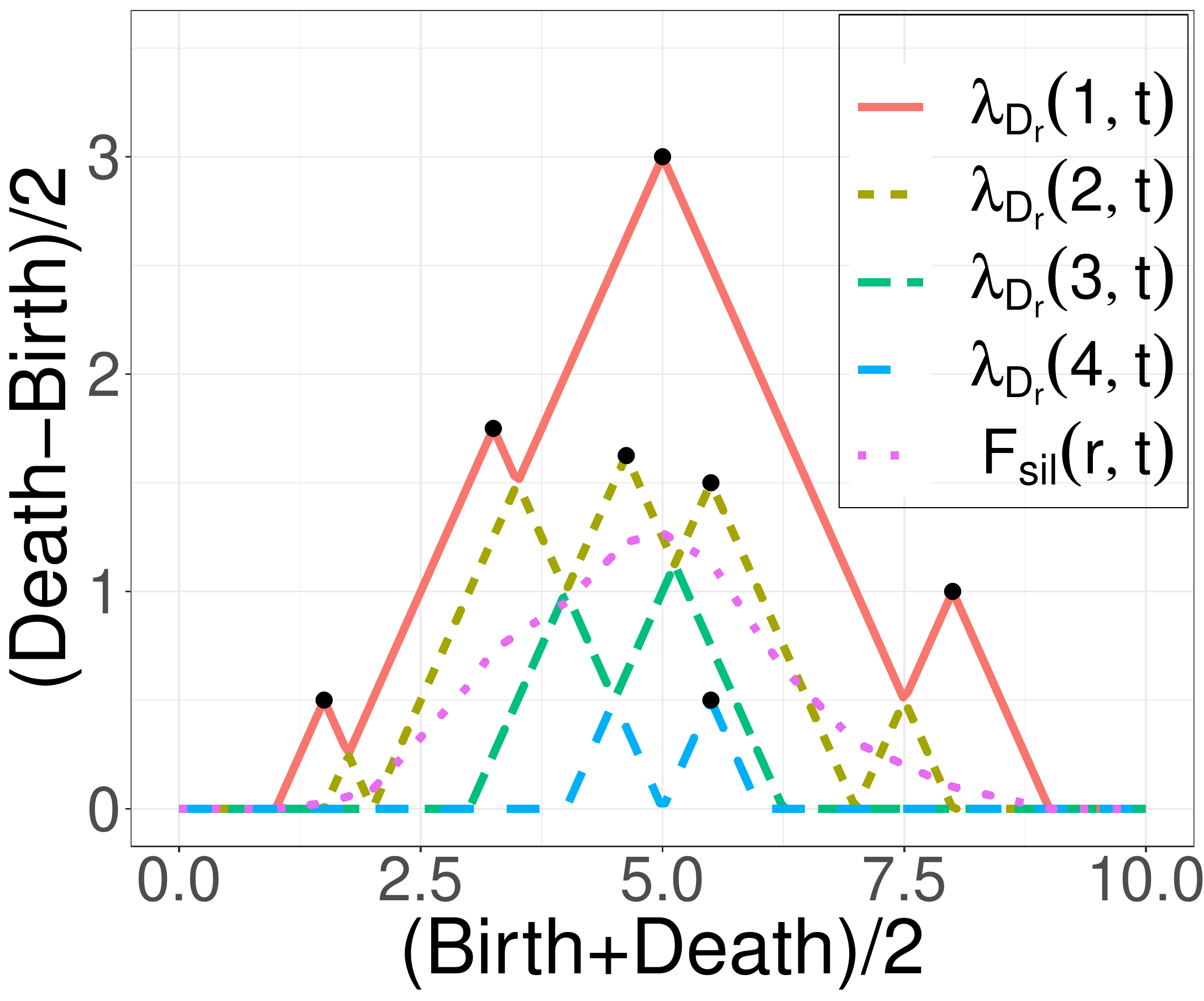}
    \caption{Summaries}
    \label{fig:landscape_fns}
  \end{subfigure}
    \caption{A persistence diagram (a) along with its landscape functions and 
    a weighted silhouette (b) for an arbitrary homology dimension $r$.  The dotted
    pink curve is the weighted silhouette function with tuning parameter 
    $p = 1$; the other four curves correspond to landscape functions $\lambda_{\diag D_r}(k, t)$ for $k = 1, \ldots, 4$.
}
    \label{fig:landscape}
\end{figure}

\paragraph{Landscape Functions}
Landscape functions \citep{bubenik2015statistical} are popular functional
summaries of persistence diagrams \citep{Bubenik:2017us, Berry:2020ws,
Bubenik:2020vq}, which are defined as follows.
Let $\diag D_r = \{(b_j,d_j)\}_{j = 1}^{n_r}$ be the finite set of
off-diagonal points of a homology dimension $r$ persistence diagram.
Next, rotate the persistence diagram such that each  point $(b_j,d_j) \in \diag D_r$ is
mapped to~$p_{r,j} =
\left(\frac{b_j+d_j}{2}, \frac{d_j-b_j}{2}\right) \in \widetilde{\diag D}_r$.
Isosceles right triangles are formed from each
$p_{r,j}$ to the base as
 \begin{equation}
 \Lambda_{p_{r,j}}(t) =
   \begin{cases}
     t - b_j  & \quad t \in [b_j, \frac{d_j+b_j}{2}]\\
     d_j - t  & \quad t \in [\frac{d_j+b_j}{2}, d_j]\\
     0  & \quad \text{ otherwise}, \\
   \end{cases}
 \end{equation}
where $t \in [t_{\min}, t_{\max}]$. The persistence landscape is then defined as
the following collection of functions
\begin{equation}
\lambda_{\diag D_r}(k, t) = \underset{p_{r,j}\in \widetilde{\diag D}_r,}{\text{kmax }} \Lambda_{p_{r,j}}(t), t \in [t_{\min}, t_{\max}], k = 1, \ldots, n_r, \label{eq:landscapes}
\end{equation}
where kmax is the $k$-th largest value.  An example of a persistence landscape
function is displayed in \figref{fig:landscape}.
Rather than working with each $k$ of $\lambda_{\diag D_r}(k, t)$ individually, a
subset of the landscape layers can be concatenated to a long vector as
\begin{equation}
\mathcal F_{\text{land}}(\mathcal I, r, t) = \oplus_{k \in \mathcal I} \lambda_{\diag D_r}(k, t),
 \label{eq:land_vec}
\end{equation}
where $\mathcal I$ is the index set of the included landscape layers.

\paragraph{Weighted Silhouette Functions}
Rather than working with each $k$ of $\lambda_{\diag D_r}(k, t)$ from
Equation~\eqref{eq:landscapes} individually, weighted silhouette functions
provide a way of combining the information in the collection of landscape
functions. Silhouettes are weighted averages of the individual functions for
homology dimension $r$ defined as
\begin{equation}
\mathcal F_{\text{sil}}(r, t \mid p) = \frac{\sum_{j = 1}^{n_r}|d_{r, j} - b_{r, j}|^p\Lambda_{p_{r,j}}(t)}{\sum_{j = 1}^{n_r}|d_{r, j} - b_{r, j}|^p}, \label{eq:sil}
\end{equation}
where the $|d_{r, j} - b_{r, j}|^p$ act as weights that can give more emphasis
or less emphasis to features with longer lifetimes depending on the
user-specified parameter $p$. The form of these weights are suggested in Ref. \cite{chazal2014stochastic}.
An example of a
weighted silhouette function is provided in \figref{fig:landscape_fns}.
More details and theoretical properties of landscapes and silhouettes can be
found in Ref. \cite{chazal2014stochastic}.

\paragraph{Euler Characteristic and Betti Functions}
The Euler characteristic (EC) is a topological invariant and can be defined as
the alternating sum of the rank of the homology groups, where the rank of the
$r$th homology group is the $r$th Betti number.
As the persistent homology filtration parameter $t$ changes and new features are
born or old ones die, the Betti numbers and EC changes, allowing for the
definition of Betti functions and an EC function.
The Betti functions can be defined as
\begin{equation}
    \mathcal F_{\text{betti}}(r, t) = |\{(r, b_j, d_j): b_j \leq t, d_j > t\}|, \label{eq:betti}
\end{equation}
which indicates the number of dimension $r$ homology group generators that
persist in the filtration at time $t$.
The only non-trivial homology groups for data in $\mathbb R^3$ are in
dimensions~0,~1, and~2; thus, the Euler characteristic equation we use is
\begin{equation}
    \mathcal F_{\text{ec}}(t) = \sum_{r=0}^2 (-1)^r \mathcal F_{\text{betti}}(r, t). \label{eq:ec}
\end{equation}
Betti and EC functions have been used in applications
\citep[e.g.,][]{mecke1993robust, park2013betti,kimura2017quantification, Pranav:2017vy,Pranav:2019ws,giri2021measuring, wilding2021persistent} and some of their theoretical
properties have been explored \citep[e.g.,][]{Bauer:2018uf, Hiraoka:2018tx,
Krebs:2019ue, Biscio:2020us, Krebs:2021vw}.

There are a number of other summary functions of persistence diagrams that have
been defined \citep[e.g.,][]{chen2015statistical, adams2015persistent,
Biscio:2019ub}.  For a general discussion of summary functions of persistence
diagrams, including some theoretical properties, see Ref. \cite{Berry:2020ws}.


\section{Methods: topological hypothesis tests for LSS} \label{sec:methods}

A primary goal in this work is to develop a framework that can inferentially discriminate between different realizations of web-like
geometric data structures such as the Cosmic Web. Our TDA-based framework allows for extracting information, encoded in large-scale
galaxy/halo distribution, that goes beyond methods commonly used in cosmology N-point clustering statistics.
The motivation is to detect differences between the DM halo spatial distributions (i.e., 3-manifolds) evolved in cosmological simulations where initial conditions were set to be either that of CDM or WDM-type.
In this section we present a hypothesis testing framework using test statistics derived from the summaries of persistence diagrams presented in \S\ref{sec:summaries}.  These topological hypothesis tests build on the work outlined in Ref. \cite{Berry:2020ws}, including their notation.


The proposed hypothesis tests rely on permutation methods to compute the p-values.  There have been some central limit theorem results for summary functions of persistence diagrams (e.g., Betti functions) and persistent homology-based hypothesis tests statistics using asymptotic theory (e.g., Refs. \cite{Yogeshwaran:2017wv, Hiraoka:2018tx, Krebs:2021vw}), but they generally assume the data were drawn from a homogeneous Poisson point process. 
Both the large and small-scale halo/galaxy distribution in the Universe cannot be described by a homogeneous Poisson process.
Owing to the nature of initial conditions (i.e., an adiabatic Gaussian random field) and the gravitational instability (a mechanism responsible
for the growth and evolution of the cosmic structures) halos spatial distribution is clustered with non-Gaussian features on small (non-linear) scales.
Naturally, also the COCO DM simulations provide halo samples which are not close to resembling homogeneous Poisson point processes, which is discussed in Appendix \S\ref{sec:poisson}.  Furthermore, as we present below, the WDM and CDM samples are not independent of one another due to the cosmological simulation design: it is therefore necessary to use matched-pairs hypothesis tests.

\subsection{Test statistic and p-value computations}
\label{sec:pvalues}

For the proposed hypothesis tests, we consider two samples of observations,
\begin{equation}
\mathbf Y_1 = \{Y_{1,1}, \ldots, Y_{1,n_1}\}, \text{ and } \mathbf Y_2 = \{Y_{2,1}, \ldots, Y_{2, n_2}\} \label{eq:data}
\end{equation}
where each $Y_{j,i}$, $i = 1, \ldots, n_j$ and $j = 1, 2$ is a data set of which a persistence diagram can be computed.  For our cosmological simulation data, each $Y_{j,i}$ will have a set of points in $\mathbb R^3$, but, in general, the  $Y_{j,i}$'s could take different forms; for example, each  $Y_{j,i}$ could be an image of a fibrin network \citep{Berry:2020ws} or a brain artery tree \citep{bendich2014persistent}.

Each observation from Equation~\eqref{eq:data} will have a corresponding persistence diagram
\begin{equation}
\diag D_1 = \{\diag D_{1,1}, \ldots, \diag D_{1,n_1}\}, \text{ and } \diag D_2 = \{\diag D_{2,1}, \ldots, \diag D_{2,n_2}\}.
\end{equation}
These samples of diagrams can be used to test the hypotheses,
\begin{equation}
H_0: \mathcal P_1 = \mathcal P_2 \text{ vs. }  H_1: \mathcal P_1 \neq \mathcal P_2, \label{eq:hypothesis_test}
\end{equation}
where $\mathcal P_1$ and $\mathcal P_2$ are the true underlying distributions of persistence diagrams from group 1 and 2, respectively.\footnote{Probability measures can be theoretically defined on a space of persistence diagrams (with a Wasserstein metric) as presented in Ref. \cite{Mileyko:2011aa}.}

Given two samples of persistence diagrams, there are a number of possible ways to derive test statistics; we consider the functional versions of persistence diagrams presented in \S\ref{sec:summaries} as test statistics.  These functional summaries can be understood as a map between the space of persistence diagrams, $\mathcal P$, to the space of functions, $\mathcal F$, defined as
$\mathbb F: \mathcal P \longrightarrow \mathcal F$.
Therefore, the diagrams from above can be used to define the collection of functional summaries with $j=1,2$ as
\begin{equation}
\diag F_j = \{F_{j,1} = \mathbb F (\diag D_{j,1}), \ldots,  F_{j, n_j} = \mathbb F (\diag D_{j, n_j})\}. \label{eq:functional_summaries}
\end{equation}

A test statistic for the two-sample hypothesis test of Equation~\eqref{eq:hypothesis_test} can be derived using estimates of functional summaries.  Letting $F_{j,i} = \mathbb F (\diag D_{j,i}), i=1, \ldots, n_j, \text{ and } j=1,2$ (see Equation~\eqref{eq:functional_summaries}), the mean functional summaries are defined as
\begin{equation}
\bar{F}_j(t)  = n_j^{-1} \sum_{i = 1}^{n_j} F_{j,i}(t). \label{eq:mean_functional_summary}
\end{equation}
Then our test statistic for the different functional summaries is based on the following distance between mean functional summaries,
\begin{equation}
d(\bar{F}_1, \bar{F}_2) = \int_{\mathbb T} | \bar{F}_1(t) - \bar{F}_2(t)| dt, \label{eq:test_statistic}
\end{equation}
where $\mathbb T$ defines the domain of the functions.
Note that the $\mathbb T$ is related to the range of values of the filtration parameter, which depends on the functional summary.  For example, for the Euler characteristic function, the $\mathbb T$ covers the range of the filtration parameter, but for the landscape and silhouette functions it represents the range of a transformed filtration parameter since the persistence diagram is rotated.

\paragraph{Matched-Pairs Permutation Test.}
Since the distributions of the test statistics of Equation~\eqref{eq:test_statistic} for the different functional summaries we consider are unknown, the p-values for the two-sample hypothesis tests can be computed using the usual permutation testing framework.  The general procedure is to randomly assign the $n_1 + n_2$ functional summaries into two groups, because this random assignment is consistent with the null hypothesis ($H_0$) from Equation~\eqref{eq:hypothesis_test} where the two groups follow the same distribution.
Using the random group assignments, the new mean functional summaries are estimated using Equation~\eqref{eq:mean_functional_summary}, $\widetilde{F}_1^{(l)} \text{ and } \widetilde{F}_2^{(l)}$, which are used to compute the distance $d(\widetilde{F}_1^{(l)}, \widetilde{F}_2^{(l)})$ from Equation~\eqref{eq:test_statistic}, for $l = 1, \ldots, n_l$ random permutations \citep{Berry:2020ws}.  The resulting (approximate) permutation p-value can then be computed as
\begin{equation}
p_{\text{perm}} = n_l^{-1}\sum_{l = 1}^{n_l} I(d(\widetilde{F}_1^{(l)}, \widetilde{F}_2^{(l)}) \geq d(\bar{F}_1, \bar{F}_2)),
\end{equation}
where $I(A)$ is an indicator function that takes the value 1 if $A$ is true and 0 if $A$ is false.

When the two sets of samples are independent, the above permutation p-value is reasonable.  However, as explained in \S\ref{sec:coco}, the CDM and WDM COCO data, $\mathbb Y_c$ and $\mathbb Y_w$, are not independent due to the initial conditions of the simulations.  Instead, the samples $\mathbf Y_{c,i}$ and $\mathbf Y_{w,i}$ for $i = 1, \ldots, 77$ have a similar spatial structure which should be accounted for in the computation of the permutation p-values.
Therefore, we consider a matched-pairs version of the permutation p-values.  The difference between this matched-pairs version and the permutation test outlined above is in how the two groups are randomly assigned for each permutation.  The matched-pairs permutation involves randomly selecting one of the two matched samples to go into each of the two groups (e.g., one of $\mathbf Y_{c,i}$ or $\mathbf Y_{w,i}$ will be randomly assigned to group 1, and the other will be assigned to group 2).
The matched-pairs permutation p-value is then defined in the same manner as above, as
\begin{equation}
p_{\text{matched}} = \sum_{l = 1}^{n_l} I(d(\widetilde{F}_{1, \text{matched}}^{(l)}, \widetilde{F}_{2, \text{matched}}^{(l)}) \geq d(\bar{F}_1, \bar{F}_2)),
\end{equation}
where $\widetilde{F}_{j, \text{matched}}^{(l)}, j = 1,2,$ are the mean functional summary for permutation $l$ using the matched-pairs random assignment.
This matched-pairs permutations p-value computation accounts for correlations between the COCO CDM and WDM samples by including one of the two matched samples within each group for each permutation, but randomizing which label (CDM or WDM) is assigned.

\section{Investigation of COCO Simulation Data} \label{sec:coco_analysis}

In order to investigate differences between the CDM  and WDM COCO samples of MW-analog halo neighborhoods described in \S\ref{sec:mw}, $\mathbb Y_c$ and $\mathbb Y_w$, respectively, we carry out the two-sample hypothesis tests defined in Equation~\eqref{eq:hypothesis_test} and described in the previous section.  The test statistics are based on the functional summaries of persistence diagrams outlined in \S\ref{sec:summaries}, along with several other methods discussed below.  The comparison methods include a test statistic that uses persistence diagrams directly (rather than a functional summary of them) and non-TDA functional summaries that capture second-order properties of spatial point processes.  The collective goals of the test statistics considered are (i) to detect differences between CDM and WDM MW-analog halo neighborhoods, and (ii) to understand and interpret any detected differences (e.g., the distance scale at which differences occur).

In addition to the test statistics using the functional summaries presented in \S\ref{sec:methods}, we also consider other approaches.  One method is the persistence diagram-based test (PDT) of Ref. \cite{Robinson:2017vm} which has a test statistic defined using distances between persistence diagrams.  We also consider two functional summaries of spatial point processes which do not use persistence diagrams, namely the G-function and the two-point correlation function (2PCF).  The G-function gives the distribution function of the nearest-neighbor distances, and 2PCF uses the Landy-Szalay estimator \citep{landy1993bias} and is one of the most basic and fundamental objects used to
study clustering in cosmology \citep{peebles1980large}.
These methods are described in more detail in Appendix \S\ref{sec:comparison}.
The p-values for these additional tests are also carried out using permutations, and we have also adapted them to work for our matched-pairs design.

\subsection{Hypothesis testing results} \label{sec:results}

All the hypothesis tests that use statistics derived from persistence diagrams (including the PDT) use the same persistence diagrams, which were computed using a VR filtration.  These computations were carried out with \emph{Ripser} \citep{Bauer2021Ripser}.
P-values were computed for the test statistics discussed previously based on 20,000 permutations using the traditional and matched-pairs permutation methods of \S\ref{sec:pvalues}.  The results are displayed in Table~\ref{tab:coco_results}.  Below we discuss the resulting p-values, and in the next section we investigate and interpret where the differences are most pronounced.

Overall, statistically significant differences with p-values $<0.001$ are apparent between the CDM and WDM MW-analog DM halo neighborhoods samples, and $p_{\text{perm}} \geq p_{\text{matched}}$ for all test statistics considered.\footnote{If our test statistics were Gaussian distributed, a p-values $<0.001$ would correspond to $> 3 \sigma$ significance.}  
Since the two sets of samples are from different populations (CDM vs. WDM COCO data), it is a positive result that our proposed tests are able to detect differences. 
For $H_0$ and $H_1$, all the function-based tests had $p_{\text{matched}} < 0.001$, and this was also the case for the G-function and 2PCF test statistics.  PDT has $p_{\text{perm}}$ and $p_{\text{matched}}\leq0.003$ for $H_0$, but higher p-values for $H_1$ with $p_{\text{perm}}=0.255$ and $p_{\text{matched}}=0.036$.  Because the PDT uses the bottleneck distance, only one $H_1$ feature on each of the persistence diagrams contribute to the test statistic for each MW-analog halo neighborhood sample, while the functional summary-based test statistics considers all the features on the persistence diagrams (except for the landscape functions which only includes features that contribute to the first 10 layers).

Aside from the silhouette function tests, the $H_2$ p-values are $<0.01$ for both $p_{\text{perm}}$ and $p_{\text{matched}}$.  The $p_{\text{perm}}$ for the silhouette function tests are $>0.10$, but then drop below $0.01$ for $p_{\text{matched}}$.
The EC function test statistic, similar to the related Betti function test statistics, has both $p_{\text{perm}}$ and $p_{\text{matched}} \leq 0.001$.
For the TDA-based test statistics, the EC, Betti, and Landscape function test statistics appear to be best able to detect differences between the CDM and WDM MW-analog halo neighborhood samples for both the traditional and matched-pairs permutation tests across the three homology dimensions ($H_0$, $H_1$, $H_2$).  Tuning could be carried out for the landscape function tests to find which landscape function layers are most informative at detecting differences.  Since the results with the first 10 layers performed well, we did not consider tuning for this analysis.

\begin{table*}
\caption{COCO data results.  Permutation p-values ($p_{\text{perm}}$) and matched permutation p-values ($p_{\text{matched}}$) for tests comparing the CDM and WDM MW-analog halo neighborhood samples.  The p-values are rounded to three decimal places and are based on 20,000 permutations as described in \S\ref{sec:pvalues}. } \label{tab:coco_results}
\begin{ruledtabular}
\begin{tabular}{ccccc}
Test statistic	&	Notation	&	Homology dimension	&	$p_{\text{perm}}$	&	$p_{\text{matched}}$	\\
\hline
Landscape	&	$\mathcal F_{\text{land}}(1:10, 0, t)$	&	0	&	0	&	0	\\
Silhouette	&	$\mathcal F_{\text{sil}}(0, t \mid p=0.5)$	&	0	&	0.009	&	0	\\
Silhouette	&	$\mathcal F_{\text{sil}}(0, t \mid p=1)$	&	0	&	0.001	&	0	\\
Silhouette	&	$\mathcal F_{\text{sil}}(0, t \mid p=2)$	&	0	&	0	&	0	\\
Betti	&	$\mathcal F_{\text{betti}}(0, t)$	&	0	&	0.001	&	0	\\
PDT	&	$\mathcal T_{\text{PDT}}(D_{1,\cdot \mid 0},D_{2,\cdot \mid 0} \mid \infty,1)$	&	0	&	0.003	&	0	\\
\hline
Landscape	&	$\mathcal F_{\text{land}}(1:10, 1, t)$	&	1	&	0	&	0	\\
Silhouette	&	$\mathcal F_{\text{sil}}(1, t \mid p=0.5)$	&	1	&	0.009	&	0	\\
Silhouette	&	$\mathcal F_{\text{sil}}(1, t \mid p=1)$	&	1	&	0.007	&	0	\\
Silhouette	&	$\mathcal F_{\text{sil}}(1, t \mid p=2)$	&	1	&	0.014	&	0	\\
Betti	&	$\mathcal F_{\text{betti}}(1, t)$	&	1	&	0	&	0	\\
PDT	&	$\mathcal T_{\text{PDT}}(D_{1,\cdot \mid 1},D_{2,\cdot \mid 1} \mid \infty,1)$	&	1	&	0.255	&	0.036	\\
\hline
Landscape	&	$\mathcal F_{\text{land}}(1:10, 2, t)$	&	2	&	0	&	0	\\
Silhouette	&	$\mathcal F_{\text{sil}}(2, t \mid p=0.5)$	&	2	&	0.123	&	0.003	\\
Silhouette	&	$\mathcal F_{\text{sil}}(2, t \mid p=1)$	&	2	&	0.158	&	0.009	\\
Silhouette	&	$\mathcal F_{\text{sil}}(2, t \mid p=2)$	&	2	&	0.135	&	0.008	\\
Betti	&	$\mathcal F_{\text{betti}}(2, t)$	&	2	&	0.001	&	0	\\
PDT	&	$\mathcal T_{\text{PDT}}(D_{1,\cdot \mid 2},D_{2,\cdot \mid 2} \mid \infty,1)$	&	2	&	0.009	&	0	\\
\hline
Euler characteristic	&	$\mathcal F_{\text{ec}}(t)$	&	0-2	&	0.001	&	0	\\
\hline
G-function	&	$\mathcal F_{\text{G}}(t)$	&	N/A	&	0	&	0	\\
2PCF	&	$\mathcal F_{\text{2PCF}}(t)$	&	N/A	&	0	&	0	\\
\end{tabular}
\end{ruledtabular}
\end{table*}

Given that we only have one COCO-CDM and one COCO-WDM realization, and that we seek to evaluate the performance of the proposed test statistics when the null hypothesis is true, we consider bootstrap samples of the data from the CDM data and from the WDM data.
The distribution of the p-values when the null hypothesis is true should follow a uniform distribution.  Details of this simulation study and the results are presented in Appendix \S\ref{sec:power}. Overall, we find the p-values resulting from proposed test statistics based on the functional summaries of persistence diagrams under the null hypothesis are generally consistent with uniform distributions.


\subsection{Interpretation of results}
\label{sec:interpretation}

In this section, we explore the Betti functions in more detail and develop other visualizations to aid in the interpretation of the results in order to investigate the scales at which the differences between the CDM and WDM MW-analog halo neighborhood samples occur and are significant.
Since our interest is in where the test statistics diverge,
the mean difference function is displayed where the signal is based on the matched data
in the CDM and WDM COCO MW-analog halo neighborhood samples using
\begin{equation}
\bar{F}_{\text{diff}}(t) = n_s^{-1} \sum_{i=1}^{n_s} \left(F_{c,i}(t) - F_{w,i}(t)\right)
\end{equation}
where $F_{c,i}(t)$ and $F_{w,i}(t)$ are functional summaries for CDM and WDM sample $i$, respectively, and $n_s=77$.  Additionally, 95\% global confidence bands are computed using the bootstrap approach outlined in Section 3.2 of Ref. \cite{Berry:2020ws}, with 1000 bootstrap samples.\footnote{Note that the hypothesis tests use $L_1$ distances between functions (see Equation~\eqref{eq:test_statistic}) while the confidence bands are investigating differences across the functions.}

\begin{figure}[htp!]
  \centering
  \begin{subfigure}{.75\linewidth}
    \includegraphics[width=\linewidth]{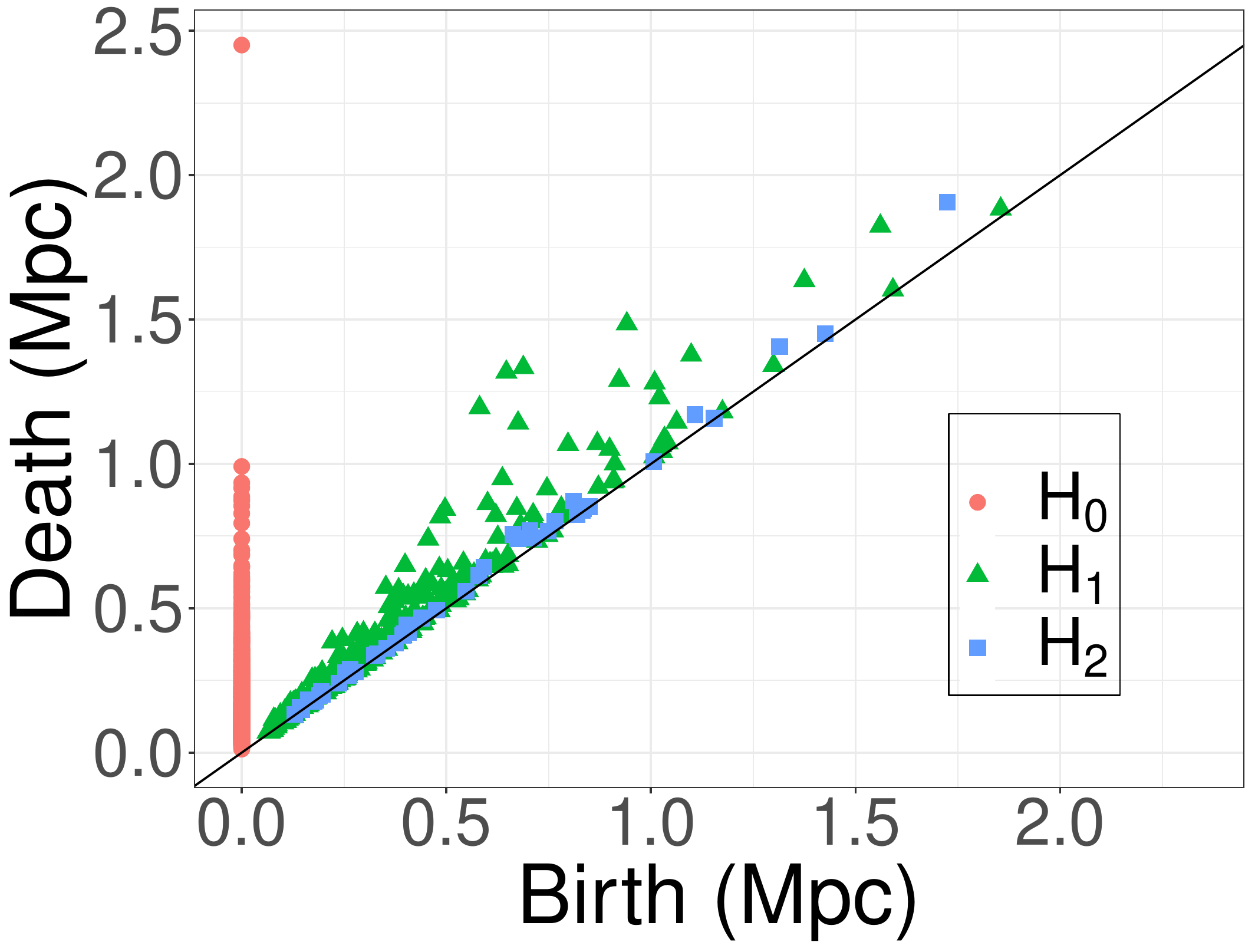}
    \caption{CDM}
    \label{fig:diag_cdm}
  \end{subfigure}
  
  \begin{subfigure}{.75\linewidth}
    \includegraphics[width=\linewidth]{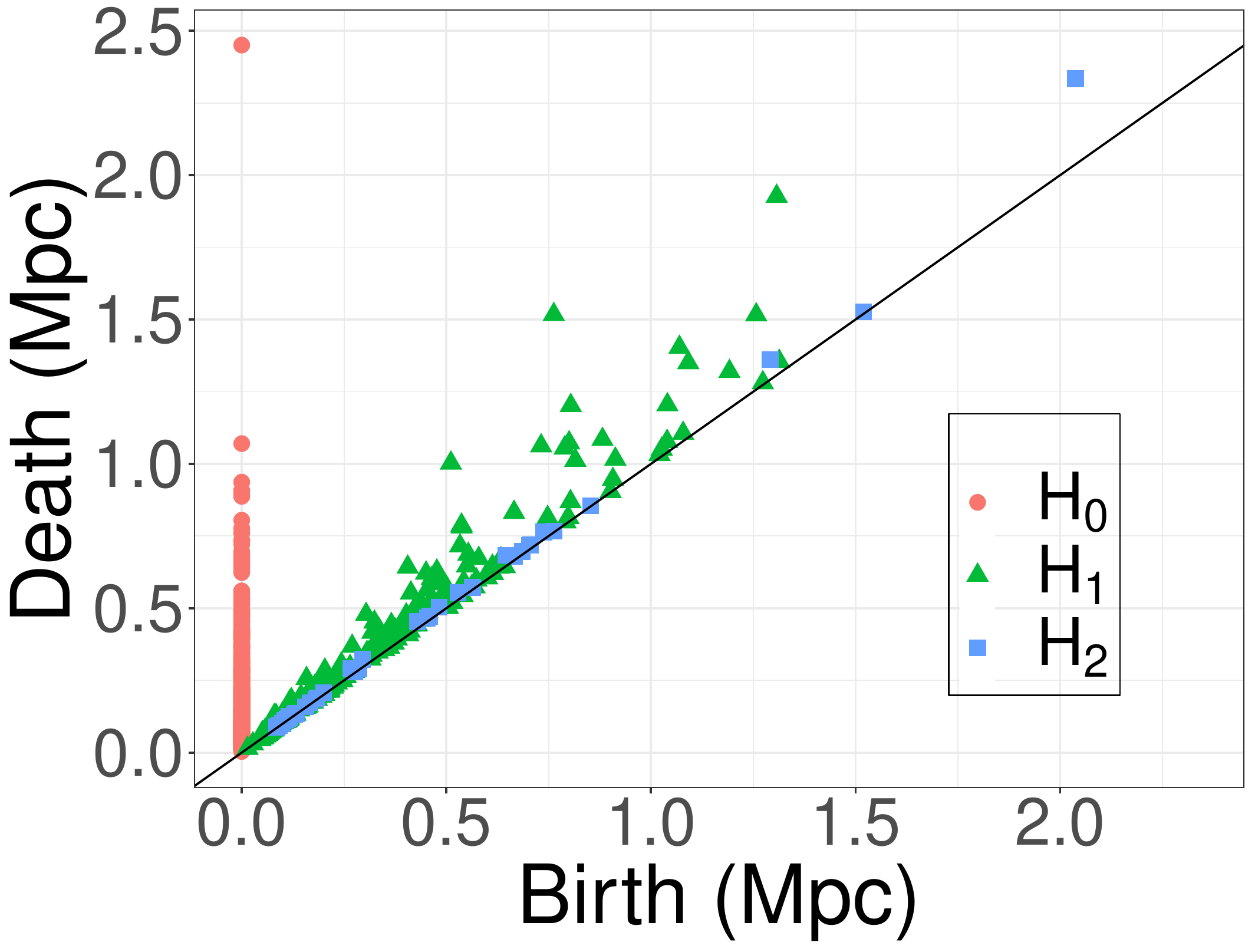}
    \caption{WDM}
    \label{fig:diag_wdm}
  \end{subfigure}
  \caption{Persistence diagrams for the MW-analog halo neighborhood sample for (a) the CDM data of \figref{fig:mw_cdm}, and (b) the WDM data of \figref{fig:mw_wdm}.
  }
  \label{fig:diag_coco}
\end{figure}

The CDM and WDM MW-analog halo neighborhood samples' persistence diagrams were generated using using a VR filtration.  For example, \figref{fig:diag_coco} displays the persistence diagrams for the COCO CDM and WDM samples of \figref{fig:mw_cdm} and \ref{fig:mw_wdm}, respectively.
The CDM and WDM persistence diagrams in this example share a similar pattern where generally the $H_0$ features are all connected by around a filtration parameter value of 1, $H_1$ features persist longer than the $H_2$ features across the range of birth times.  The $H_0$ feature plotted on both diagrams at (0, 2.57) represents an $H_0$ feature that in fact persists indefinitely and should, technically, be plotted at a death time of infinity.

\begin{figure*}[htp!]
  \centering
  \begin{subfigure}{0.3\textwidth}
    \includegraphics[width=\linewidth]{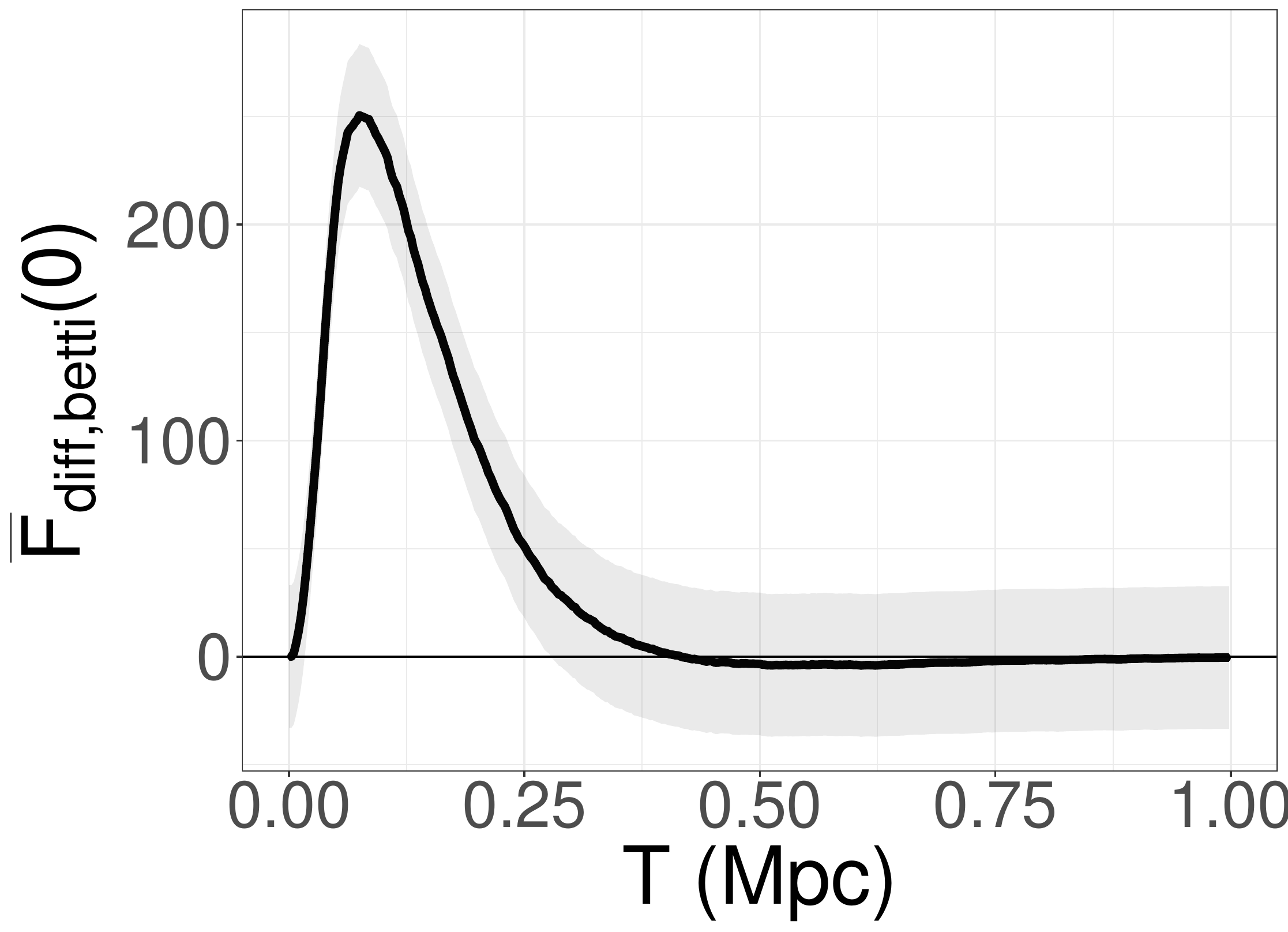}
    \caption{$\mathcal F_{\text{betti}}(0, t)$}
    \label{fig:h0_betti}
  \end{subfigure}
  \begin{subfigure}{0.3\textwidth}
    \includegraphics[width=\linewidth]{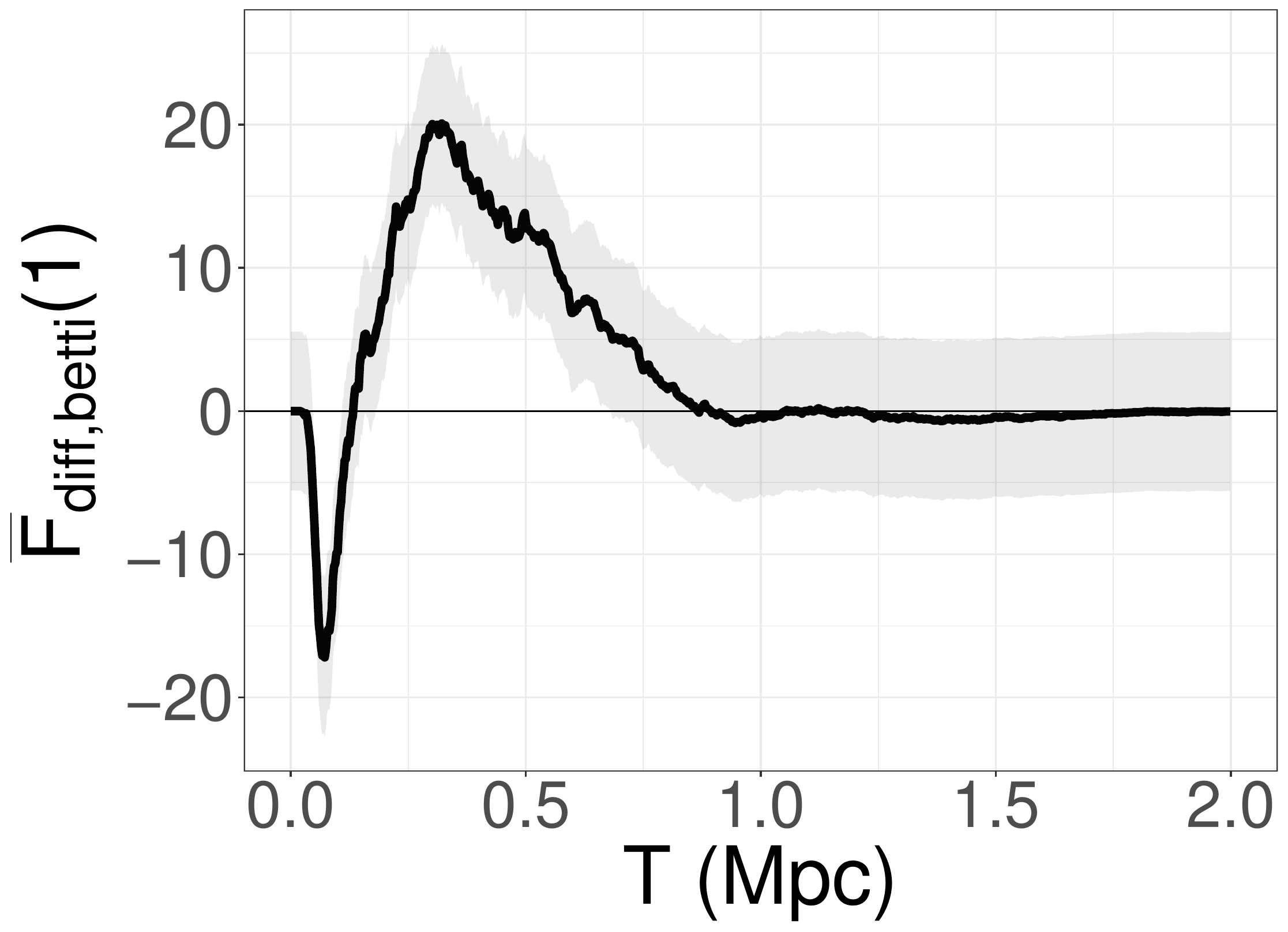}
    \caption{$\mathcal F_{\text{betti}}(1, t)$}
    \label{fig:h1_betti}
  \end{subfigure}
  \begin{subfigure}{0.3\textwidth}
    \includegraphics[width=\linewidth]{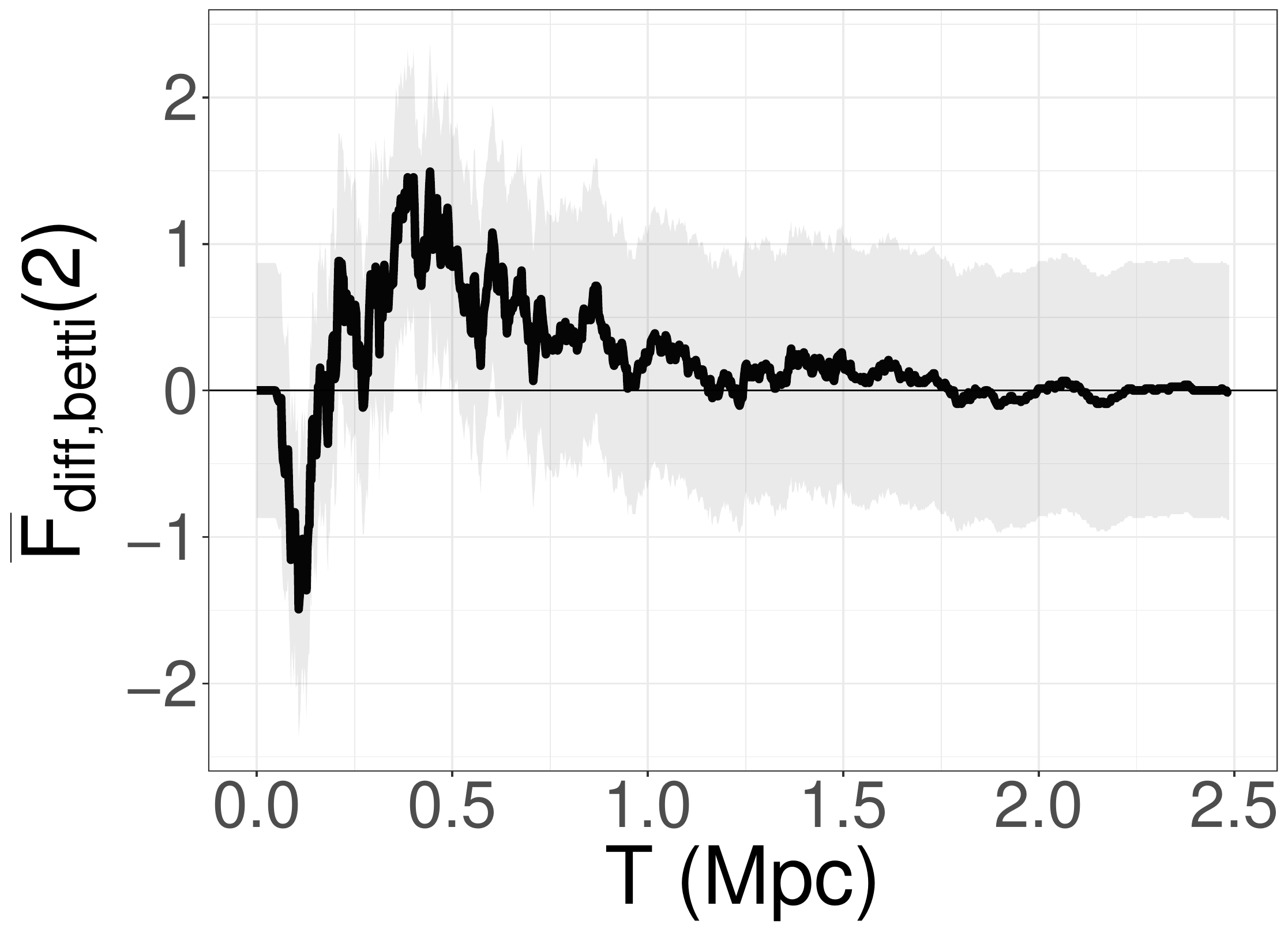}
    \caption{$\mathcal F_{\text{betti}}(2, t)$}
    \label{fig:h2_betti}
  \end{subfigure}
  \caption{Mean differences (CDM-WDM) of the Betti functional summaries along with 95\% confidence bands (shaded regions) for the noted functional summaries. The x-axis limits were set to highlight the non-zero mean differences regions.
  } \label{fig:functions}
\end{figure*}

The mean differences (WDM - CDM) of the Betti functional summaries are displayed in \figref{fig:functions}  along with the corresponding 
95\% confidence bands.
Overall, these summaries suggest that the CDM and WDM samples differ on shorter distance scales, but then start to resemble each other at 
longer distance scales in keeping with \figref{fig:introData}.  Recall that the Betti functions count the number of features that are persistent at 
the filtration parameter values (i.e., the x-axis) so by considering the average difference of the Betti functions we observe at which scales the number of features differ between the CDM and WDM.
For $H_0$, the number of features, on average, for the CDM data is larger than the number for the WDM for distances until scales of around 0.4 Mpc, 
and then the number of WDM features is slightly higher than the number of CDM features until distances of  $\sim$0.75 Mpc.
The number of $H_1$ features is greater, on average, for the WDM data over the CDM data when $t\leq 0.13$ Mpc, and 
then the CDM has more $H_1$ features until around 0.9 Mpc.  A similar pattern is observed with the $H_2$, but the average differences between 
the CDM and WDM are within only two $H_0$ features.

\begin{figure}[htp!]
  \centering
  \begin{subfigure}{0.23\textwidth}
    \includegraphics[width=\linewidth]{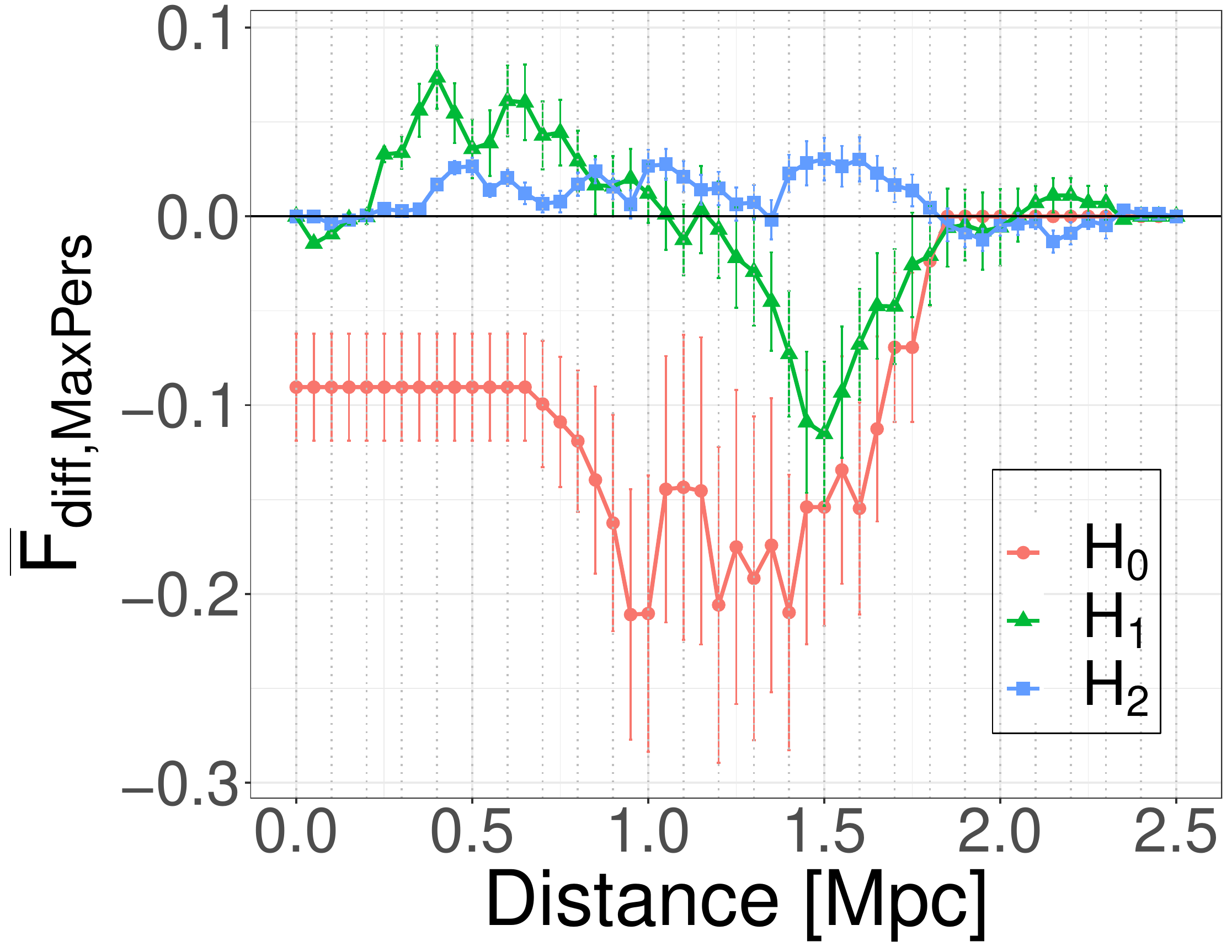}
    \caption{Max persistence}
    \label{fig:maxper}
  \end{subfigure}
  \begin{subfigure}{0.23\textwidth}
    \includegraphics[width=\linewidth]{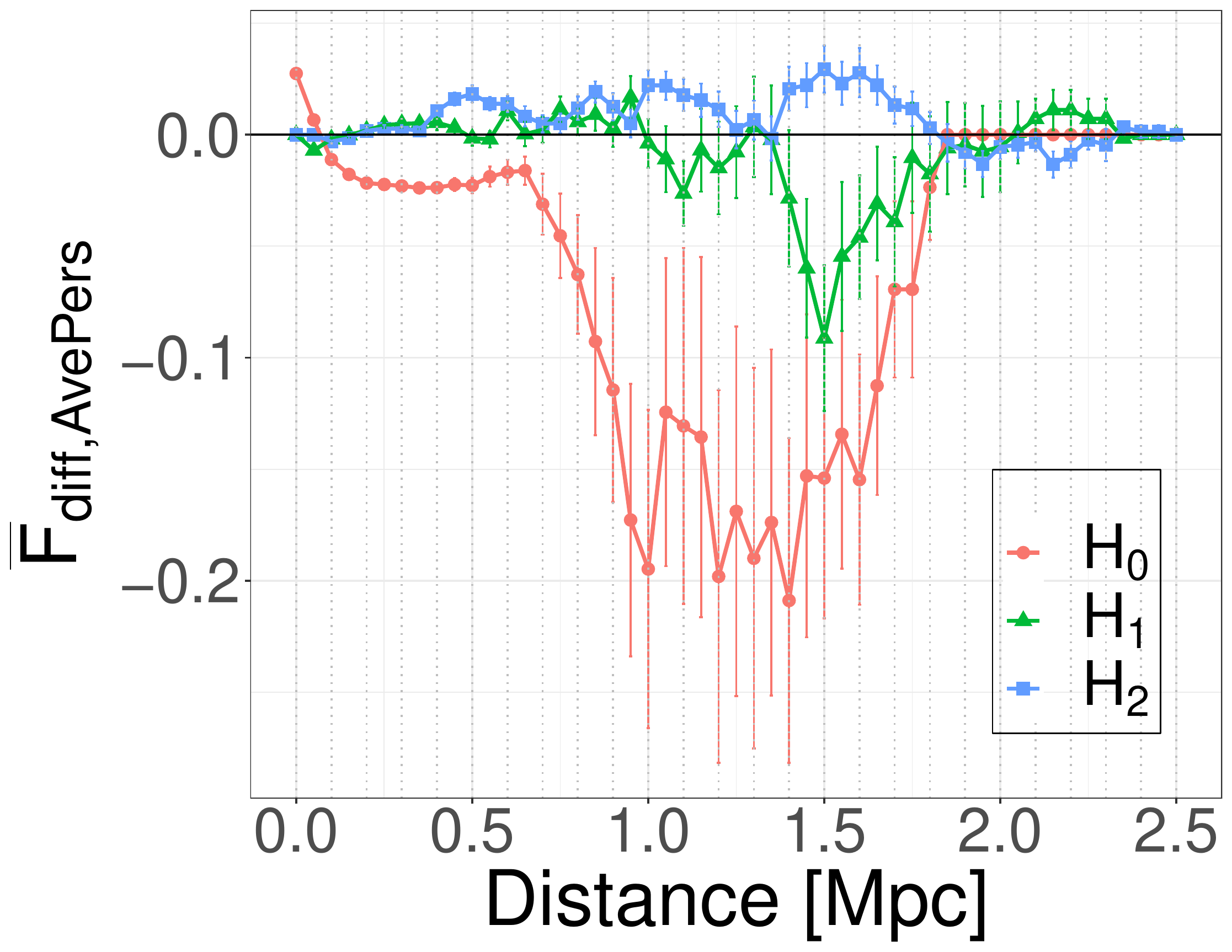}
    \caption{Average persistence}
    \label{fig:aveper}
  \end{subfigure}
  \caption{Mean differences (CDM-WDM) of the maximum (a) and average (b) persistences $\pm$ one standard error. Means and standard errors were computed every 0.05 Mpc between 0 and 2.5. Gray dotted vertical lines are plotted every 0.10 Mpc. Note that the $H_0$ feature that persists indefinitely has been removed from this analysis.
  } \label{fig:functions_persistence}
\end{figure}

While Betti functions capture the number of features that persist across the filtration parameter values, we defined analogous functions that instead capture the maximum persistence (MaxPers) and average persistence (AvePers), which are displayed in \figref{fig:maxper} and \figref{fig:aveper}, respectively.  Similar to the plots in \figref{fig:functions}, the mean difference (CDM-WDM) of these MaxPers and AvePers functions for the matched samples were computed.  However, for \figref{fig:functions_persistence}, in order to visualize the variability in the mean differences, pointwise error bars ($\pm$ one standard error) are included. The filtration parameter grid ranges from 0 to 2.5 Mpc with a spacing of 0.05. This is a lower resolution than the Betti function figures, which we adopt here in order to be able to improve the visibility of the individual error bars.
There are larger differences between CDM and WDM MaxPers in $H_0$, $H_1$, and $H_2$
 for $t\simlt1.85$ Mpc: generally the $H_0$ MaxPers are greater for WDM than CDM, the $H_1$ MaxPers is greater 
for CDM than WDM at scales $\simlt$1.1 Mpc when this tendency switches and WDM has greater MaxPers, and the $H_2$ MaxPers are higher for CDM than WDM.
A similar pattern is apparent with the AvePers functions except the $H_1$ AvePers are similar for CDM and WDM until scales 
around 1 Mpc, after which WDM generally has greater AvePers until around 2 Mpc.

\begin{figure}[htp!]
  \centering
  \begin{subfigure}{0.23\textwidth}
    \includegraphics[width=\linewidth]{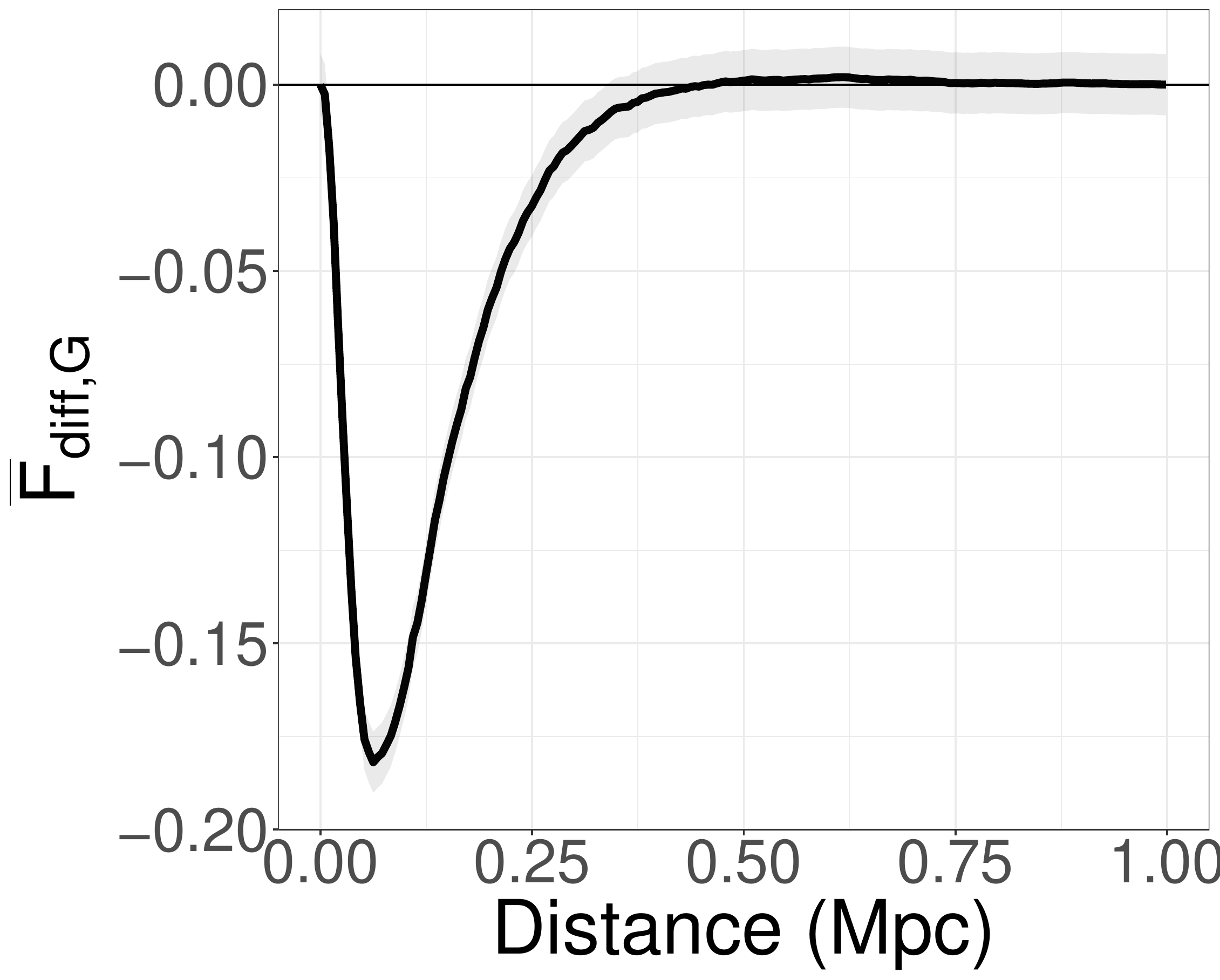}
    \caption{$\mathcal F_{\text{G}}(t)$}
    \label{fig:g}
  \end{subfigure}
  \begin{subfigure}{0.23\textwidth}
    \includegraphics[width=\linewidth]{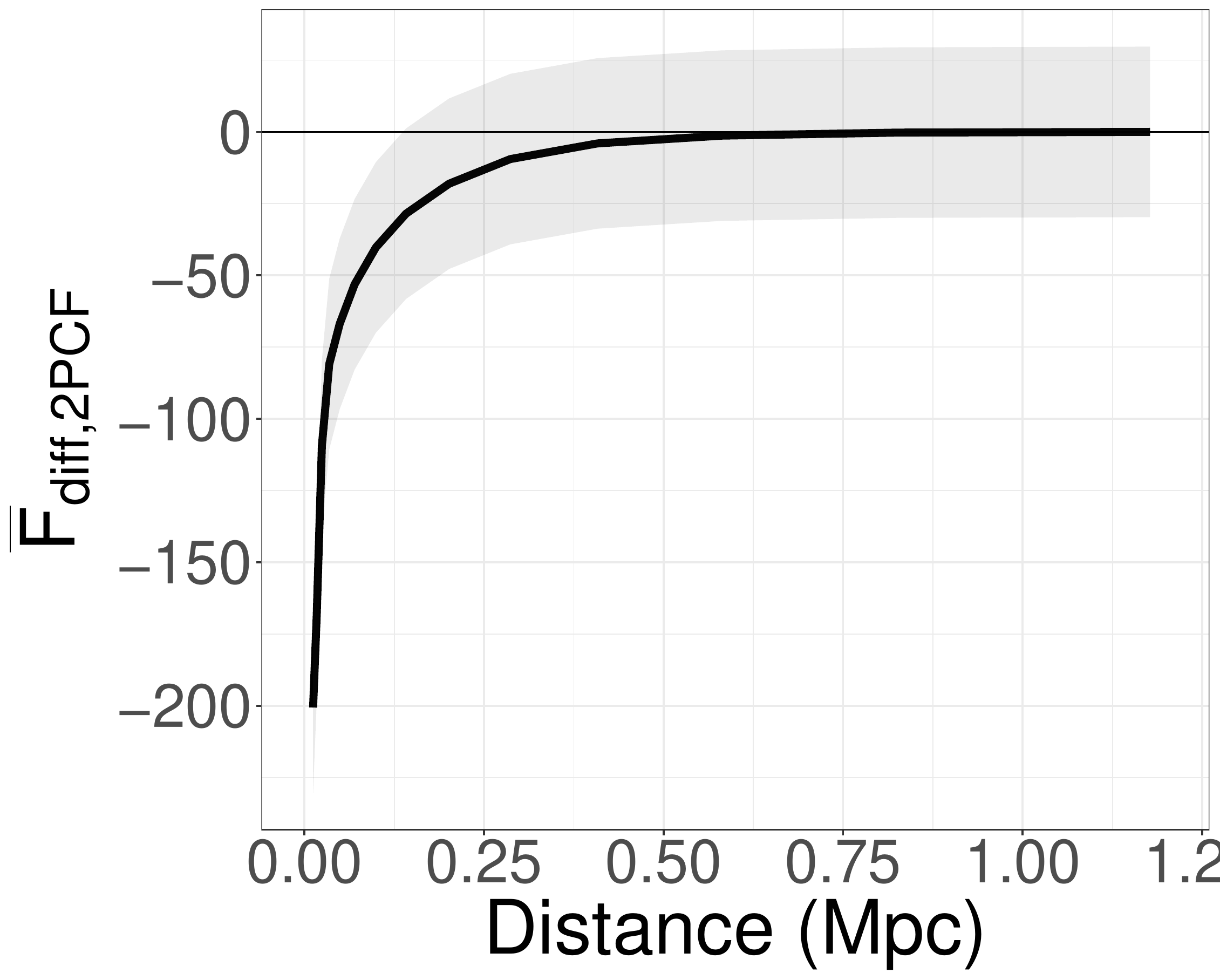}
    \caption{$\mathcal F_{\text{pcf}}(t)$}
    \label{fig:pcf}
  \end{subfigure}
  \caption{Mean differences (CDM-WDM) of the point process functions along with 95\% confidence bands (shaded regions).
  } \label{fig:pp}
\end{figure}

Basic spatial point process summary functions, such as the 2PCF, are commonly employed tools in cosmological large-scale structure study.
To quantify the degree to which our persistence diagrams provide new information over these standard statistics, we calculate and
show mean difference functions for the G-functions and 2PCFs in \figref{fig:g} and \figref{fig:pcf}, respectively.
The plotted data indicate that the WDM functions take, on average, greater values than the CDM variants for $t\simlt 0.5$ Mpc.  This result points toward a similar direction as what we observed with the differences in the $H_0$ Betti function mean differences displayed in \figref{fig:h0_betti}.  This is not surprising since the $H_0$ Betti functions, the G-functions, and the 2PCFs have different ways of assessing the closeness of the halos within the samples. However, the $H_1$ and $H_2$ Betti functions, together with the MaxPers and AvePers  methods, appear to detect differences between the CDM and WDM at different scales, suggesting that they provide distinct information from the spatial point process functions.
In particular, the $H_1$ and $H_2$ functions suggest that as the halos become connected (i.e., the death of $H_0$ features), the CDM and WDM models are forming loops and voids (i.e., $H_1$ and $H_2$ features, respectively) in different ways.  Also, the MaxPers and AvePers of the $H_0$ features differ between the CDM and WDM data on different scales than those of the $H_0$ Betti functions.

\section{Conclusion}
\label{sec:conc}

The LSS contains valuable information about the composition, and evolution, and the physical nature of the Universe.
TDA tools such as persistent homology provide a novel opportunity to extract this information from cosmological data.
While TDA-based approaches have been applied in various fields of statistical studies, its application
to cosmological data and analysis is still in its infancy.
In this paper, we introduced a hypothesis testing framework built on persistent homology that extends the work of Ref. \cite{Berry:2020ws} in order to compare topological summaries of MW-analog halo neighborhoods (3~Mpc spheres) evolved under two different DM models: CDM and WDM (WDM thermal relic mass: 3.3~keV). Next, we have assessed the sensitivity and robustness of this framework in the context of differentiating between
CDM and WDM variants.
The proposed collection of test statistics based on persistence diagrams uses summaries that were recently proposed in the literature \citep{chazal2014stochastic, bubenik2015statistical, Bubenik:2017us, Pranav:2017vy, Berry:2020ws,Bubenik:2020vq}, and are easier to work with than the original persistence diagrams. The results of the persistence diagram-based functional summaries were compared to two spatial point process functional summaries (G-functions and 2PCF) and test statistics that use persistence diagrams directly (PDT) \citep{Robinson:2017vm}.

We showed empirically that such a framework is able to infer differences between CDM and WDM, and investigated the scales at which differences occur. While most of the test statistics were able to detect statistically significant differences with $p_{\text{matched}}\leq 0.009$ for all tests considered except the PDT for $H_1$  (\S\ref{sec:results}, especially Table~\ref{tab:coco_results}), the persistent homology-based functional summaries appear to detect differences between the CDM and WDM data on different scales from the spatial point process functional summaries (\S\ref{sec:interpretation}).

Our results imply that the homology properties of clustered CDM and WDM haloes distributions are very different on small scales. CDM haloes are distributed across a larger number of clusters (homology dimension~0) than WDM haloes, especially at the $\sim80$~kpc filtration scale (Fig.~\ref{fig:h0_betti}), although the clusters that form in WDM are more persistent on average (Fig.~\ref{fig:aveper}). The $80$~kpc scale is also where the two process functions return the biggest difference between the 
models---in both cases an excess of clustering in CDM relative to WDM---plus the filtration scale at which WDM features more loops (homology dimension~1) than CDM. We thus build a picture in which CDM rapidly builds up a large number of small clusters, whereas WDM builds a smaller number of clusters, many of which will be rapidly converted into loops. 

This picture is consistent with the formation of haloes in and around cosmological filaments. In CDM, the distribution of filaments extends to near arbitrarily small scales and fills much of configuration space, whereas the WDM cutoff restricts WDM haloes to lie along large filaments and so their spatial distribution is much more constrained. Therefore, the dispersed CDM haloes form large numbers of small, isolated clusters, whereas WDM haloes are quickly joined up along cosmological filaments into loops. 

The question remains as to whether this difference between the models can be detected in the spatial distribution of observed Local Group galaxies. One will have to select haloes that are likely to form a galaxy, where most of the haloes that we included in this study will be below the HI cooling limit and thus dark \cite{Benson02,Sawala16}. Reducing the number of haloes available in this manner will likely lead to a reduction in the statistical significance of differences between the models' persistent homology properties: it is therefore imperative to make halo selections based on, for example, peak halo mass or a semi-analytic models \cite{Bose17} to confirm the potential persistent homology has for understanding which DM models best describe the Local Group.  


\appendix

\section{Distributional Assumptions of DM Samples}
\label{sec:poisson}

In this section, we carry out tests to show that the spatial distributions of the halo samples do not follow a homogeneous Poisson point process (i.e., \emph{complete spatial randomness}, CSR), which then precludes the use of many theoretical results that rely on that assumption; see \S\ref{sec:methods} for a brief discussion about some asymptotic results in persistent homology.
Ref. \cite{Baddeley:2015ve} describe a straightforward Monte Carlo test for checking CSR.  Using the same number of observations (i.e., the number of halos in the MW-analog halo neighborhoods) and the same window volume (i.e., a sphere with radius 3 Mpc), $N_{\text{MC}}$ Monte Carlo realizations are generated assuming CSR and then their G-functions are estimated; see Appendix \S\ref{sec:comparison} and Equation~\eqref{eq:g} for background on G-functions.  Then a global envelope is defined using the $N_{\text{MC}}$ summary functions based on the maximum absolute deviation of the simulated summary functions from the (known) theoretical summary function (assuming CSR).
If the summary function for the observations are outside the band, then that is evidence against CSR for those data.

\begin{figure}[htp!]
  \centering
\includegraphics[width=0.75\linewidth]{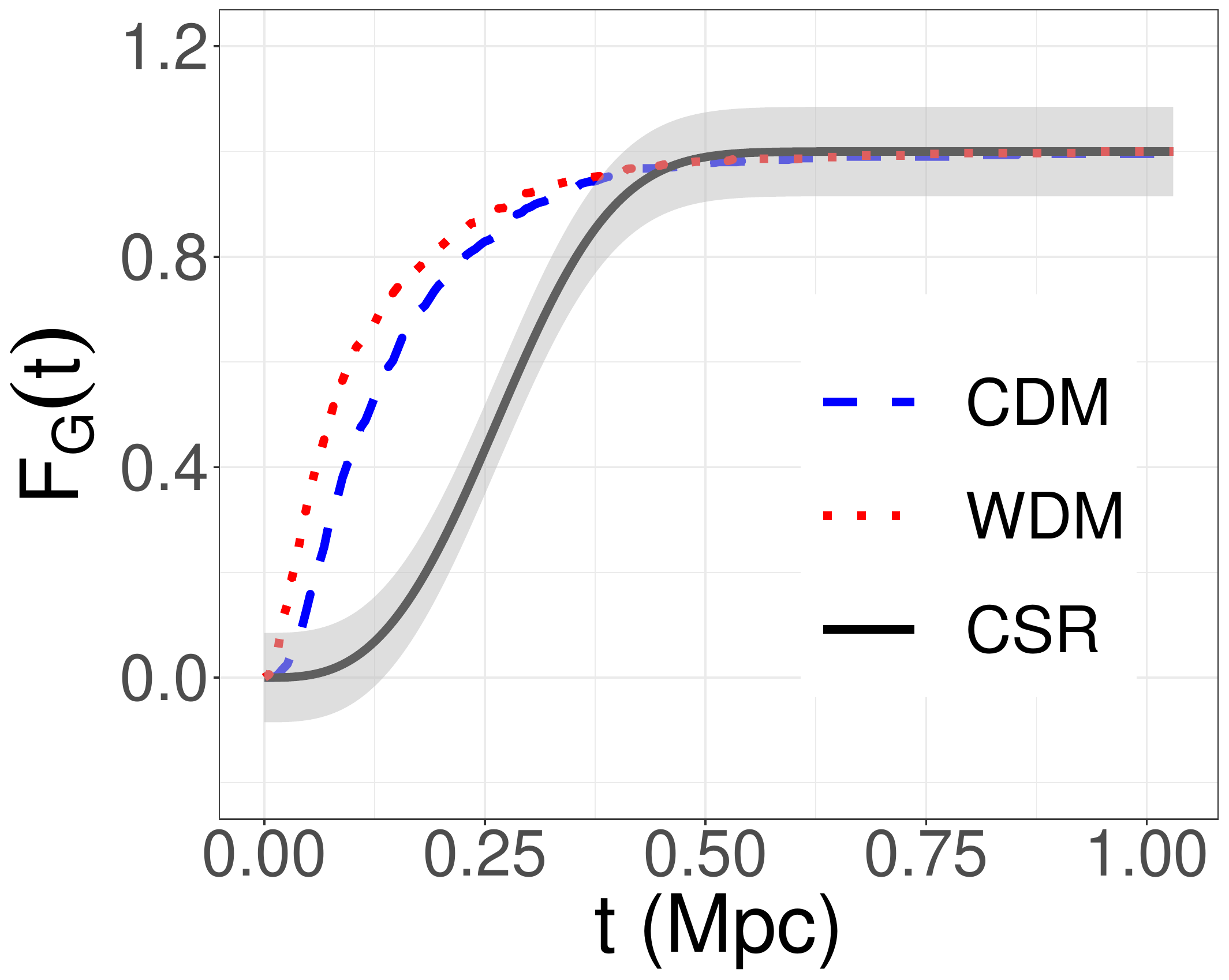}
  \caption{The estimated spatial point process summary functions for the CDM (dashed blue line) and WDM (dotted red line) MW-analog halo neighborhood samples displayed in \figref{fig:mw_example}, along with the theoretical summary function assuming CSR (solid black line) and its global envelope (gray region) using $N_{\text{MC}}=19$ samples.
  Because this is a global envelope, we can reject the hypothesis that the CDM and WDM samples were generated from homogenous Poisson point process at the $1/(1+N_{\text{MC}}) = 0.05$ level of significance.
  } \label{fig:pp_csr}
\end{figure}

The global envelope was computed by generating $N_{\text{MC}}$ realizations of a homogeneous Poisson process within a sphere of radius 3 Mpc using rejection sampling.  The number of points was set to match the number of halos in each of the MW-analog halo neighborhood samples.  Then a G-function was estimated for each sample using the \emph{G3est} function in the \emph{spatstat} R package.
For each simulated G-function, the maximum absolute deviation was computed using the true G-function of the corresponding Poisson process, defined as
\begin{equation}
    \mathcal F_G(t) = 1 - e^{-\frac{4}{3}\pi\hat{\lambda} t^3}
\end{equation}
where $\hat{\lambda}$ is the intensity estimated as the number of points divided by the volume of the sphere.
The interpretation is that if the observed G-function is outside the envelope for any value $t$ Mpc, then we can reject CSR at a significance level of $1/(1+N_{\text{MC}})$ (Ch. 10, Ref. \cite{Baddeley:2015ve}).
Using $N_{\text{MC}} = 19$, these global envelopes were computed
for all 77 samples of the CDM and WDM data, and all observed G-functions have regions outside the envelopes.
For illustration purposes, the resulting global envelope and observed G-functions for the CDM and WDM MW-analog halo neighborhood samples from \figref{fig:mw_example} are displayed below in \figref{fig:pp_csr}.

\section{Comparison Methods in COCO Analysis} \label{sec:comparison}

In addition to the test statistics proposed based on functional summaries of persistence diagrams, we include three comparison test statistics in our investigation of the COCO simulation data presented in \S\ref{sec:coco_analysis}.  The comparison methods are the Persistence Diagram Test (PDT), and test statistics derived using the G-function and two-point correlation function (2PCF) which are popular functional summaries of spatial point processes.  The comparison methods are described below.

\paragraph{Persistence Diagram Test (PDT)}
Ref. \cite{Robinson:2017vm} developed a two-sample test that compares persistence
diagrams rather than functional summaries of persistence diagrams.
The PDT test statistic takes the following form,
\begin{eqnarray}
    \mathcal T_{\text{PDT}}&&(D_{1,1:n_1 \mid r},D_{2,1:n_2 \mid r} \mid p,q) =  \nonumber \\
    &&\sum_{l=1}^2 \frac{1}{2n_l(n_l-1)} \sum_{i=1}^{n_l} \sum_{j=1}^{n_l} W_p(D_{l,i\mid r}, D_{l,j\mid r})^q 
\end{eqnarray}
where $D_{l,1:n_l\mid r}$ is a set of $n_l$ persistence diagrams for homology
dimension $r$ from population~$l = 1, 2$, $q$ satisfies $1 \leq q < \infty$,
and $W_p(\cdot, \cdot)$ is the
$p$-Wasserstein distance, with~$1 \leq p \leq \infty$.
In this work, we set $q=1$ and $p=\infty$. The $W_{\infty}$ distance is
also known as the \emph{bottleneck distance}, and is
defined as
\begin{equation}
    W_{\infty}(D_1, D_2) = \inf_{\eta: D_1 \rightarrow D_2} \sup_{x \in D_1} \|x - \eta(x)  \|_{\infty} \label{eq:bottleneck}
\end{equation}
where $D_1$ and $D_2$ are persistence diagrams,
$\eta$ defines a bijection between the two persistence diagrams that allows
for matches to the diagonal $\Delta$,
and
$\|\cdot\|_{\infty}$ is the $L_{\infty}$ norm in $\R^2$
computed between the birth and death coordinates of $x$ and $\eta(x)$ for a fixed homology dimension $r$.

\paragraph{Spatial Point Process Functions}
In order to investigate properties of the spatial distributions of the data, which in our setting is the location of the DM halos, we consider a popular functional summary of spatial point processes, the G-function\footnote{Also referred to as the ``nearest-neighbor distance distribution function"}, along with the 2PCF which is commonly used in cosmology research.
The G-function is defined below and estimated using the implementation for three-dimensional point patterns in the R package {\tt spatstat} \citep{Baddeley:2004wi, Baddeley:2014wq};\footnote{The R function from the {\tt spatstat} package is {\tt G3est}.
}
see Ref. \cite{Baddeley:2015ve} for more details. The 2PCF uses the Landy-Szalay estimator \citep{landy1993bias}

The G-function and 2PCF are estimated for each sample of the DM halos, $\mathbf Y_{k,i} \in \mathbb R^{n_i \times 3}$ for $k = w,c$ and $i = 1, \ldots, 77$.
Given a sample $\mathbf Y \in \mathbb R^{n \times 3}$, let each point be denoted by $Y_i = (Y_{i,1}, Y_{i,2}, Y_{i,3})$ for $i = 1, \ldots, n$.
Define a distance function, $\rho$, as
\begin{equation}
\rho(x,\mathbf A) = \inf\{\|x - a\| : a \in \mathbf A \}
\end{equation}
which represents the shortest distance between some point $x \in \mathbb R^3$ and a closed set $A \subset \mathbb R^3$.
The G-function gives the distribution function of the nearest neighbor distances, and can be defined as
\begin{equation}
\mathcal F_G(t) =  \mathbb P(\rho(Y_i,\mathbf Y_{-Y_i}) \leq t \mid Y_i \in \mathbf Y) \label{eq:g}
\end{equation}
where $\mathbf Y_{-Y_i}$ is the set of points $\mathbf Y$ excluding the point $Y_i$.  The Kaplan-Meier estimator of Ref. \cite{Baddeley:1997tz} is used to address the edge effects (i.e., boundary issues).


\section{Distribution of p-Values Under the Null Hypothesis}
\label{sec:power}

\begin{figure}[htp!]
  \centering
    \begin{subfigure}{.23\textwidth}
        \centering
        \includegraphics[width=\linewidth]{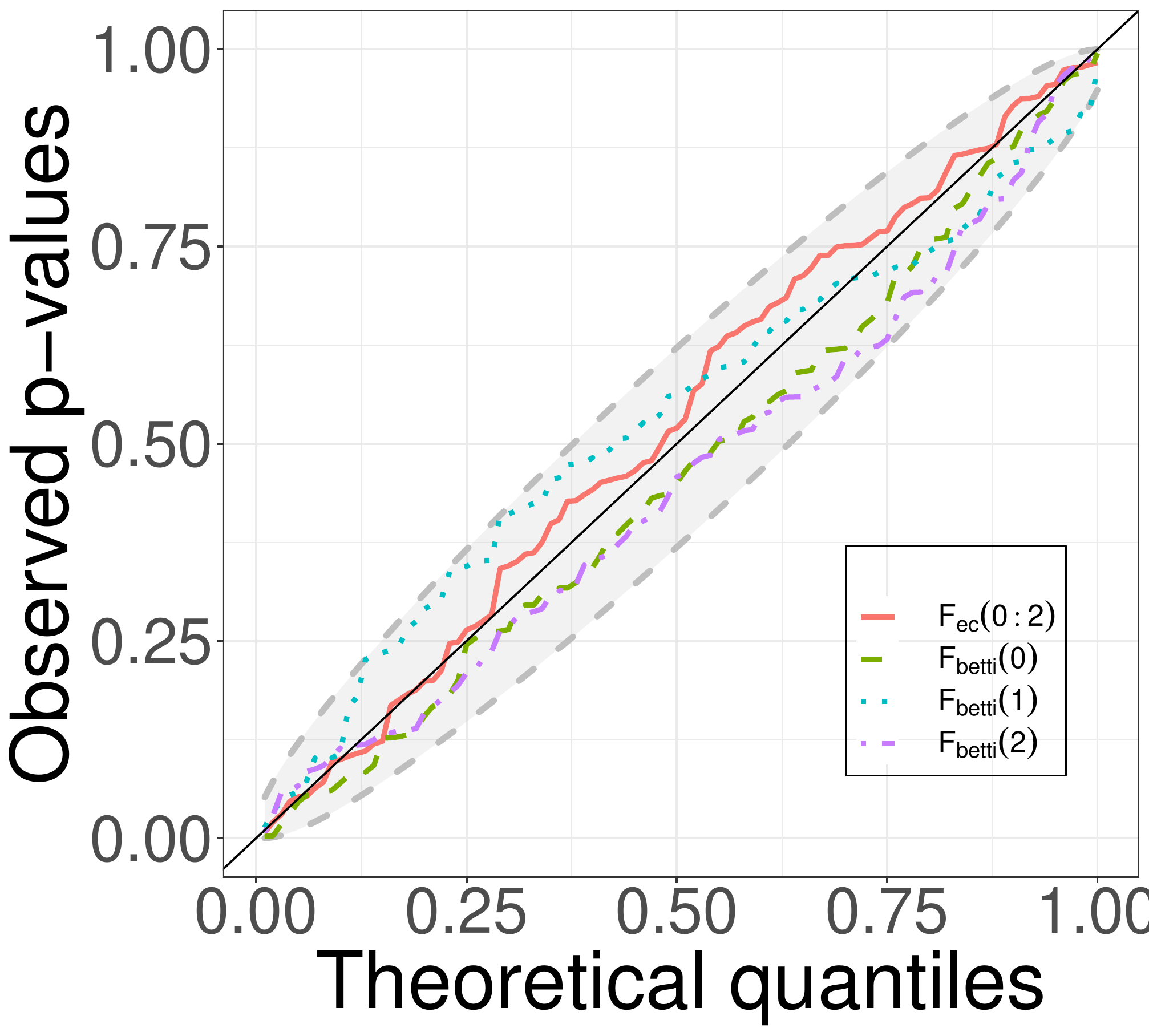}
        \caption{Betti functions}
        \label{fig:power_cdm_ec}
    \end{subfigure}
    \begin{subfigure}{.23\textwidth}
        \centering
        \includegraphics[width=\linewidth]{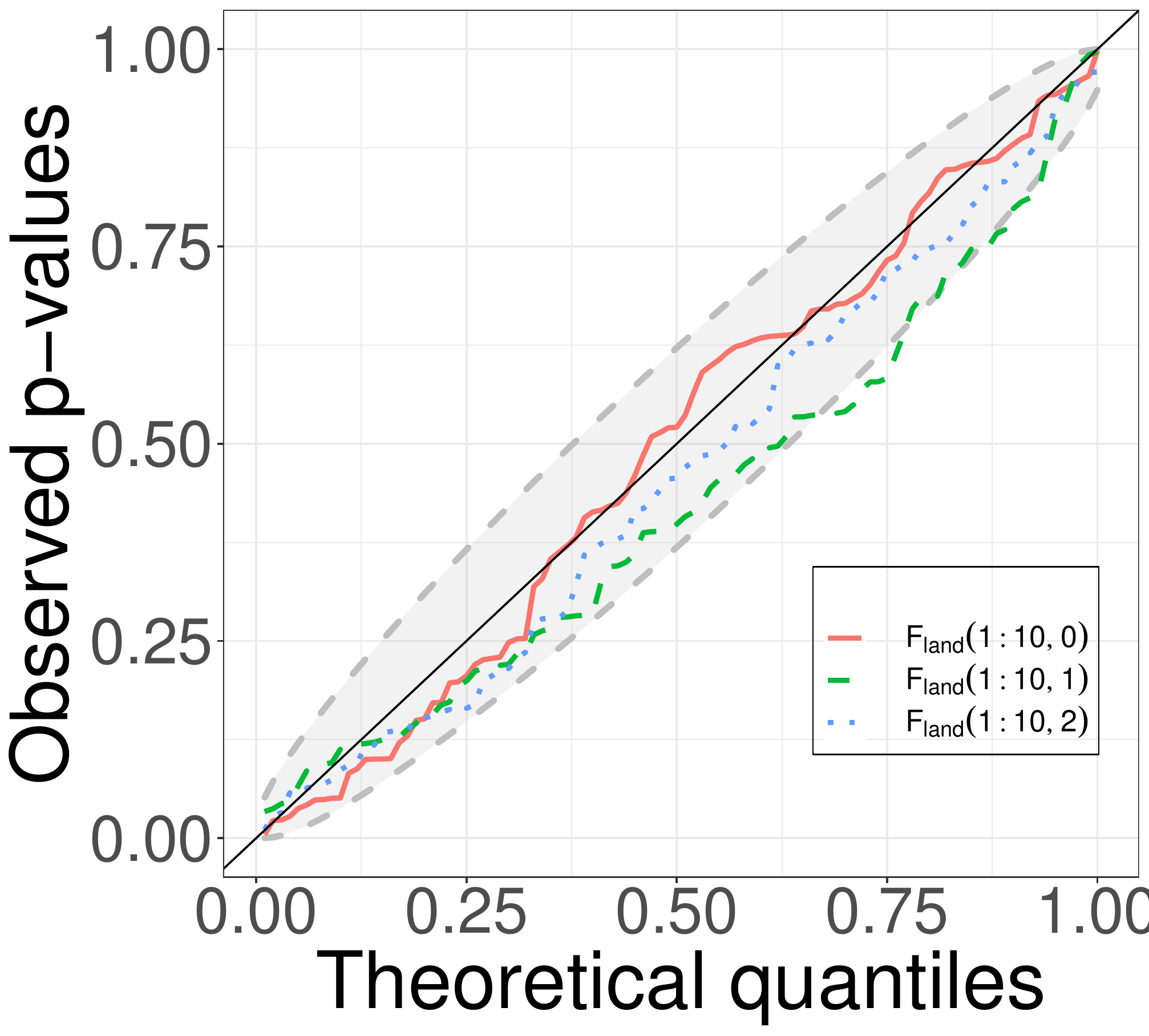}
        \caption{Landscape functions}
        \label{fig:power_cdm_land}
    \end{subfigure}\\
    \begin{subfigure}{.23\textwidth}
        \centering
        \includegraphics[width=\linewidth]{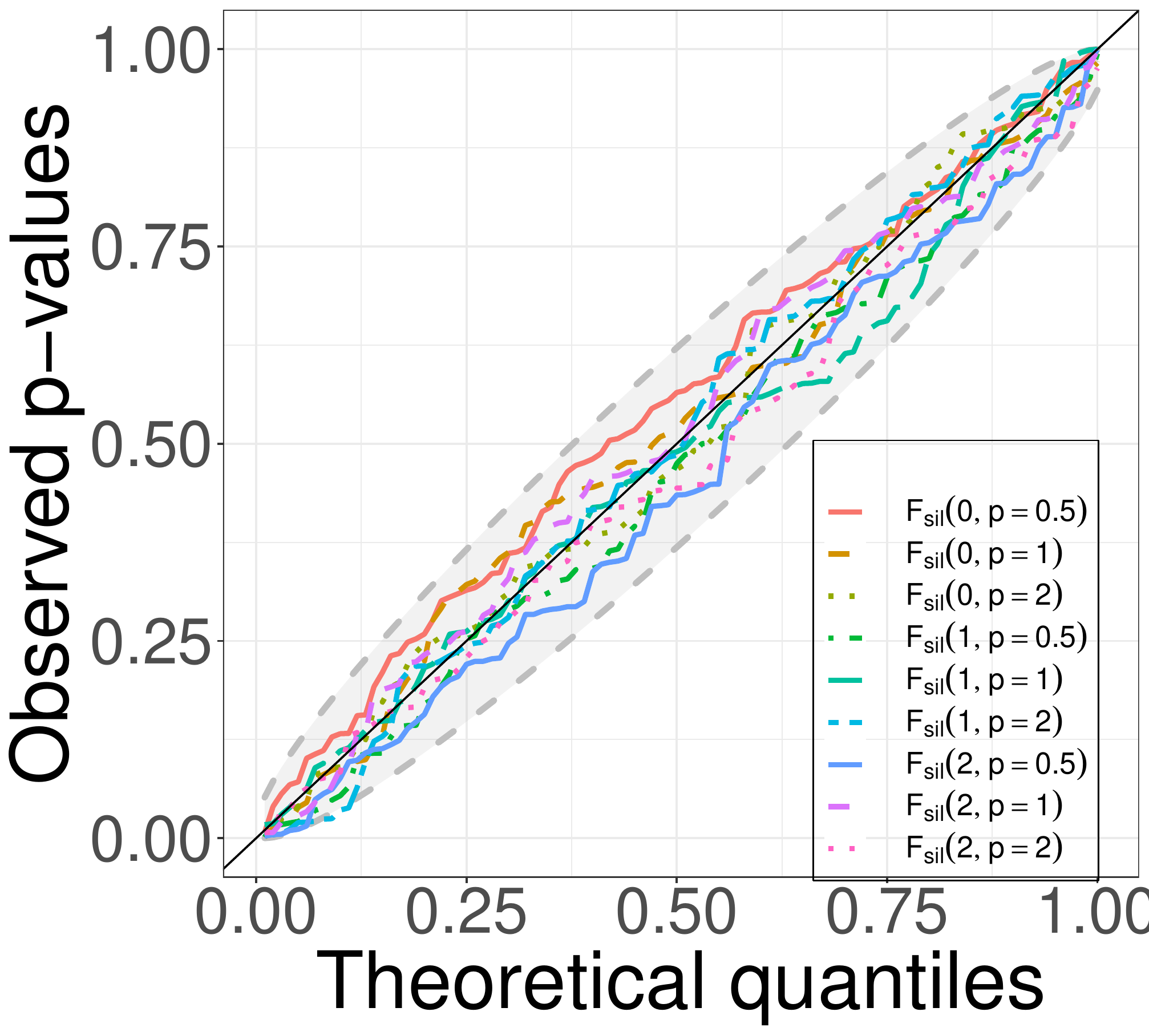}
        \caption{Silhouette functions}
        \label{fig:power_cdm_sil}
    \end{subfigure}
    \begin{subfigure}{.23\textwidth}
        \centering
        \includegraphics[width=\linewidth]{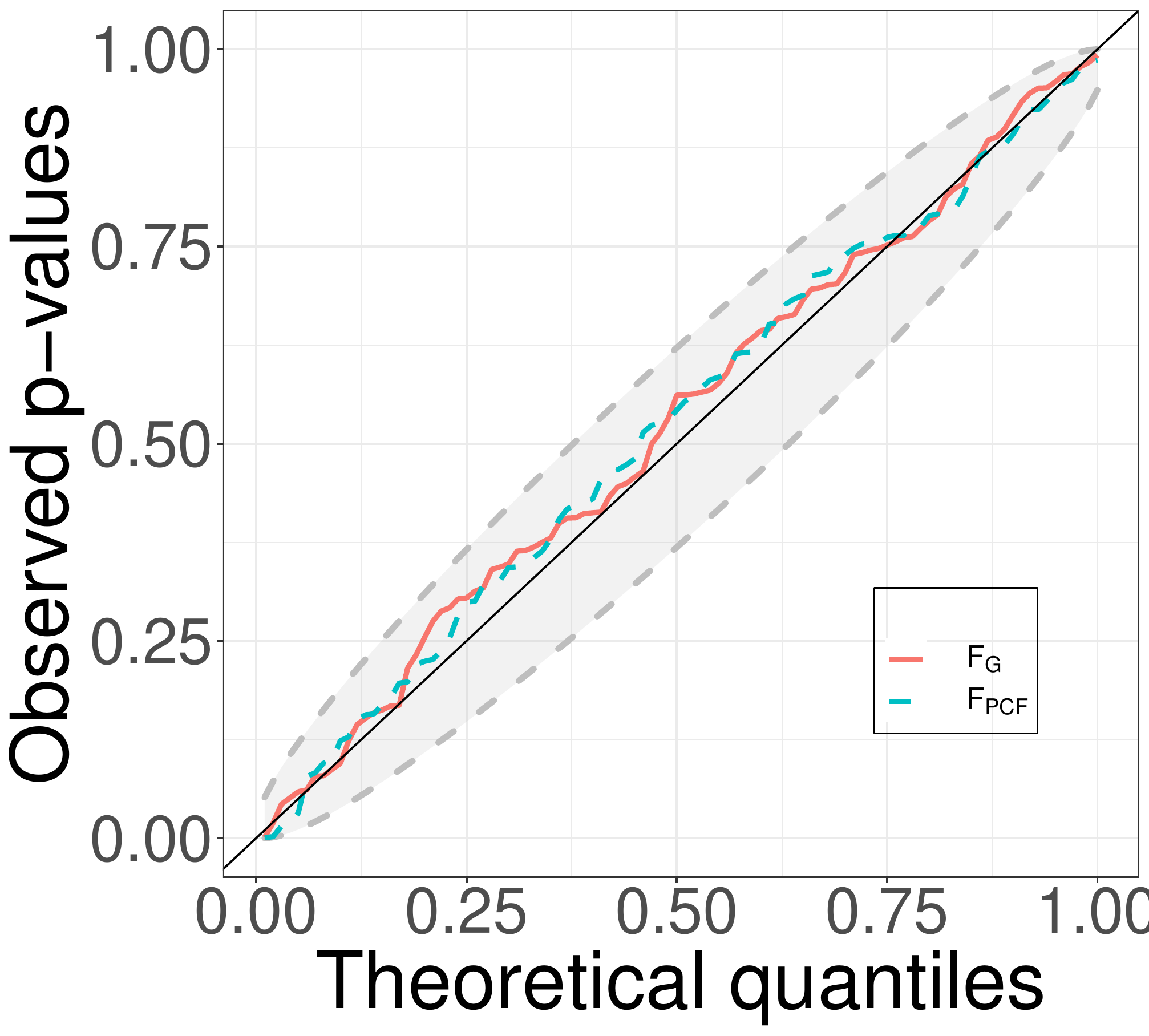}
        \caption{Spatial functions}
        \label{fig:power_cdm_spatial}
    \end{subfigure}
    \caption{Uniform quantile-quantile plots of the 100 permutations p-values calculated for each test statistic using bootstrap realizations of the CDM MW-analog halo samples.  Each bootstrap sample includes 77 MW-analog halo neighborhoods, and 20,000 permutations were used to compute each p-value.}
    \label{fig:power_cdm}
\end{figure}

\begin{figure}[htp!]
  \centering
    \begin{subfigure}{.23\textwidth}
        \centering
        \includegraphics[width=\linewidth]{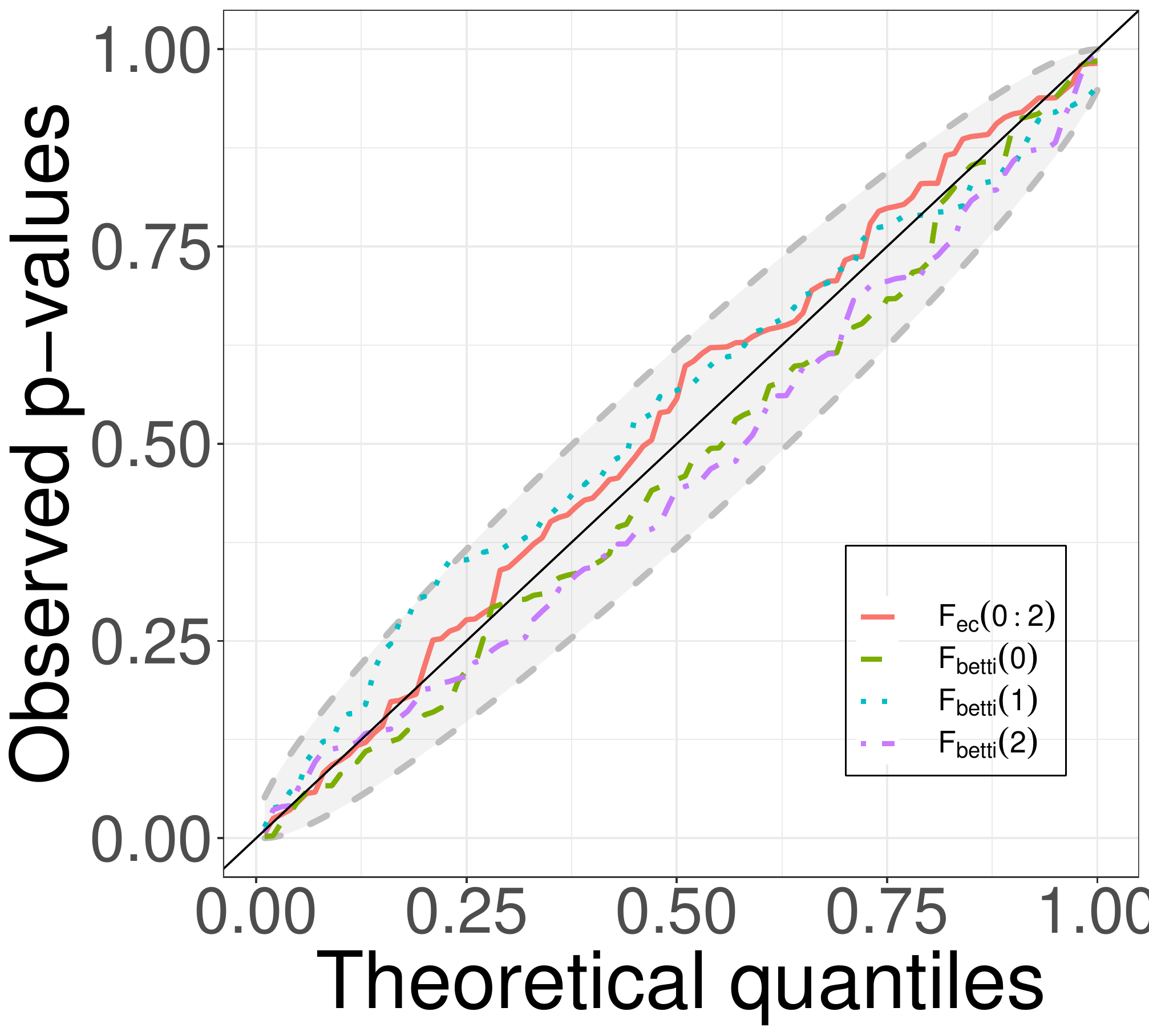}
        \caption{Betti functions}
        \label{fig:power_wdm_ec}
    \end{subfigure}
    \begin{subfigure}{.23\textwidth}
        \centering
        \includegraphics[width=\linewidth]{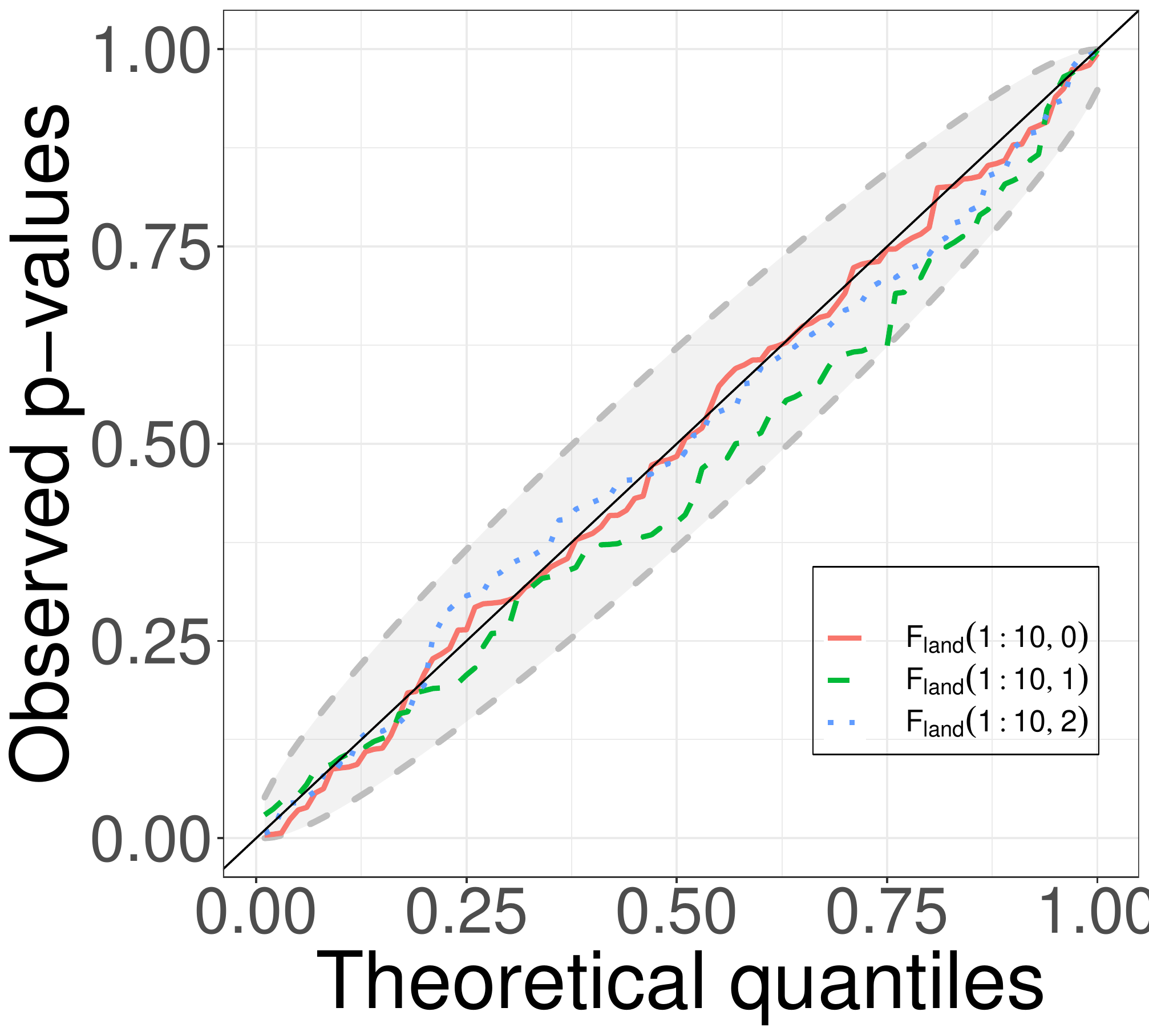}
        \caption{Landscape functions}
        \label{fig:power_wdm_land}
    \end{subfigure}\\
    \begin{subfigure}{.23\textwidth}
        \centering
        \includegraphics[width=\linewidth]{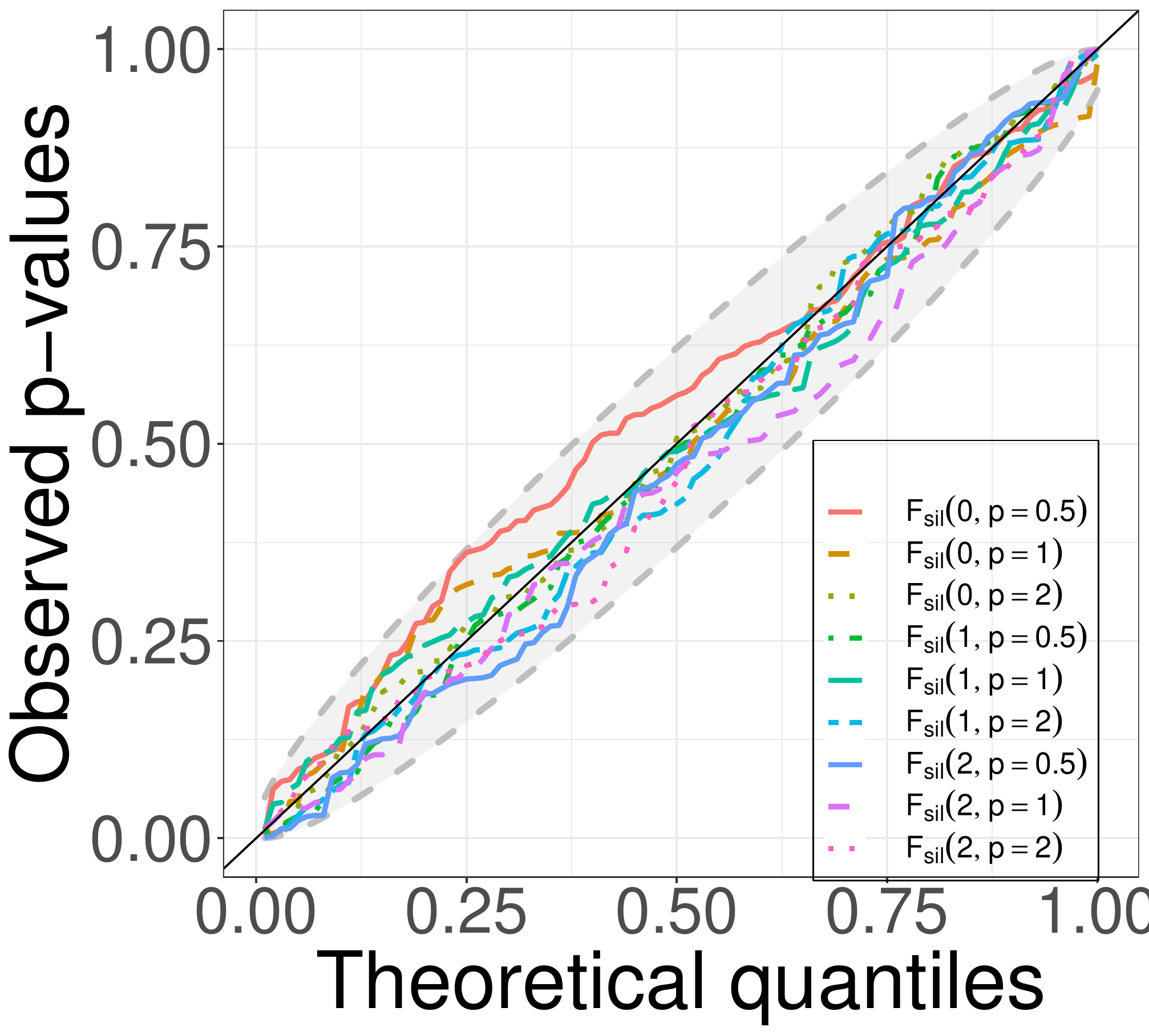}
        \caption{Silhouette functions}
        \label{fig:power_wdm_sil}
    \end{subfigure}
    \begin{subfigure}{.23\textwidth}
        \centering
        \includegraphics[width=\linewidth]{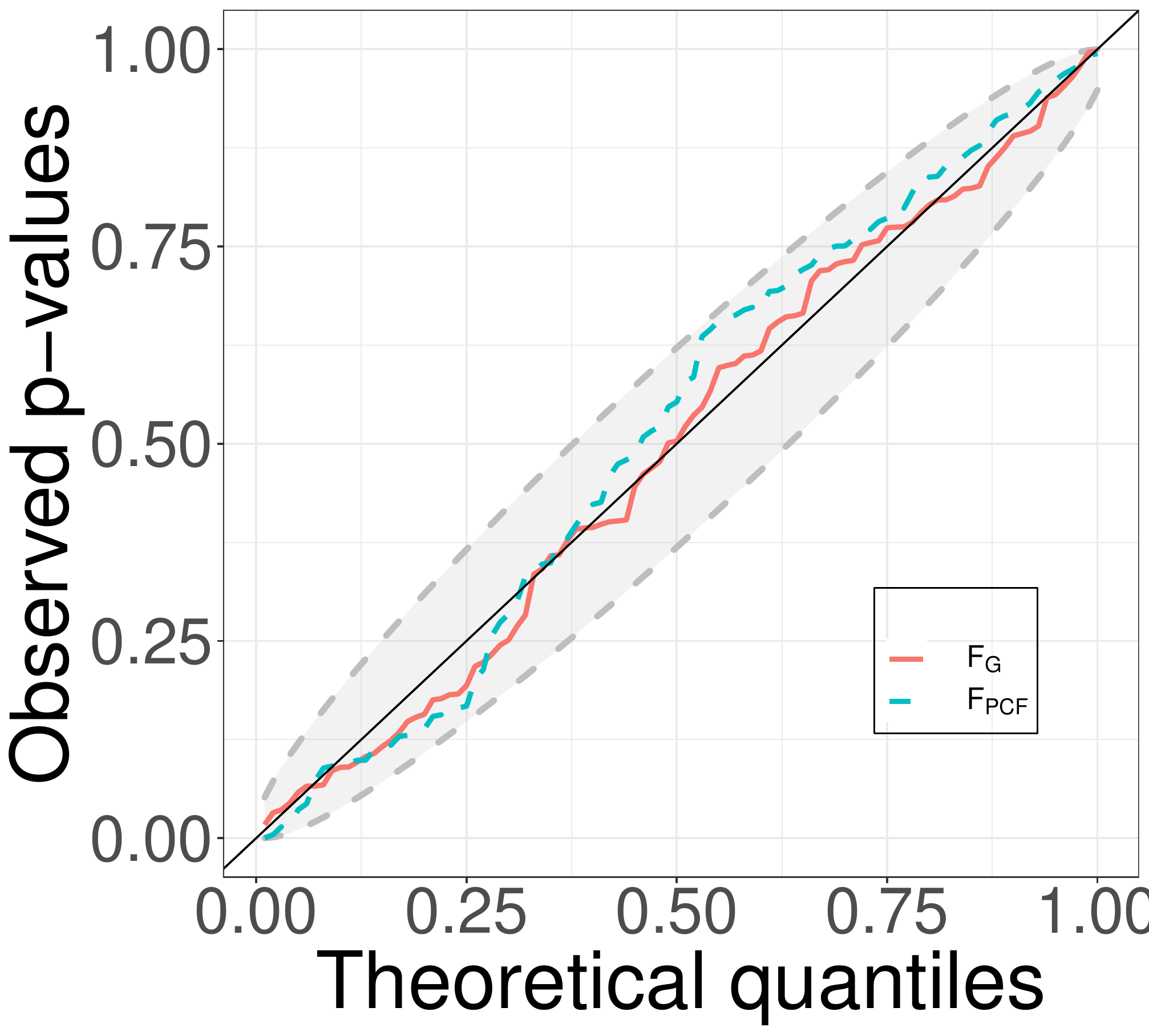}
        \caption{Spatial functions}
        \label{fig:power_wdm_spatial}
    \end{subfigure}
    \caption{Uniform quantile-quantile plots of the 100 permutations p-values calculated for each test statistic using bootstrap realizations of the WDM MW-analog halo samples.  Each bootstrap sample includes 77 MW-analog halo neighborhoods, and 20,000 permutations were used to compute each p-value.}
    \label{fig:power_wdm}
\end{figure}

The results of the proposed hypothesis tests are presented in \S\ref{sec:results}.  Many of the test statistics find statistically significant differences between the CDM and WDM models with p-values $\leq 0.001$.  In order to verify that the test statistics do not inappropriately reject the null hypothesis when the null hypothesis is true (i.e., when both groups come from either CDM or WDM), we carry out the following experiment.
We repeatedly generate two sets of boostrap realizations from either the CDM or WDM samples, and then compute permutation p-values for the test statistics presented in the main text.
The distribution of the p-values in this setting where the null hypothesis is true should follow a uniform distribution.
To compute one p-value, two bootstrap samples (with replacement) of 77 MW-analog halo neighborhoods are selected from the CDM (WDM) data.  Then the hypothesis testing framework presented in \S\ref{sec:pvalues} is used to compute a traditional permutation p-value (since the matched pairs design is not present in this setting) using 20,000 permutations.  This computation is repeated for 100 independent iterations for the CDM (WDM) data with the same sampled indexes used for the CDM and WDM bootstrap samples.
\figref{fig:power_cdm} and \figref{fig:power_wdm} display the results for the CDM and WDM samples, respectively, as uniform quantile-quantile plots with 99\% pointwise bands based on the distribution of order statistics of uniform random variables (i.e., Beta($k$, $n+1-k$) where $k$ is the order and $n=100$). The resulting p-values for each test statistic are generally consistent with uniform distributions.  The CDM $H_1$ landscape function p-values (\figref{fig:power_cdm_land}) have some values that are not within the 99\% confidence band, but this does not occur with the  WDM $H_1$ landscape function p-values (\figref{fig:power_wdm_land}) nor with the other landscape function p-values so it appears to not be a reason for concern about the landscape function-based test statistics.

\begin{acknowledgments}
This research was performed using the compute resources and assistance of the UW-Madison Center For High Throughput Computing (CHTC) in the Department of Computer Sciences. The CHTC is supported by UW-Madison, the Advanced Computing Initiative, the Wisconsin Alumni Research Foundation, the Wisconsin Institutes for Discovery, and the National Science Foundation, and is an active member of the Open Science Grid, which is supported by the National Science Foundation and the U.S. Department of Energy's Office of Science.
    This work used the DiRAC Data Centric system at Durham University, operated by the Institute for Computational Cosmology on behalf of the STFC DiRAC HPC Facility (www.dirac.ac.uk). This equipment was funded by BIS National E-infrastructure capital grant ST/K00042X/1, STFC capital grants ST/H008519/1 and ST/K00087X/1, STFC DiRAC Operations grant ST/K003267/1 and Durham University. DiRAC is part of the National E-Infrastructure.
    This project has also benefited from numerical computations performed at the Interdisciplinary Center for Mathematical and Computational Modeling (ICM) University of Warsaw under grants \#no GB79-7, GA67-17 and G63-3. 
    JCK and BTF acknowledge support from NSF under Grant Numbers DMS 2038556 and 1854336.
    WAH and PD acknowledge the support from the Polish National Science Center within research projects no.  2018/31/G/ST9/03388, 2020/39/B/ST9/03494.
   MRL acknowledges support by a Grant of Excellence from the Icelandic Research Fund (grant number 206930).
\end{acknowledgments}

\bibliography{mybib}

\end{document}